\documentclass[a4paper,10pt,pra,twocolumn]{revtex4}
\usepackage[dvips]{graphics,color}
\usepackage{amsmath}
\usepackage{amsfonts}

\usepackage{amssymb}
\usepackage{amstext}
\usepackage[sort&compress]{natbib}
\begin{document}
\newcommand{\id}{{\sf 1 \hspace{-0.3ex} \rule{0.1ex}{1.52ex}\rule[-.01ex]{0.3ex}{0.1ex}}}
\newcommand{\ignore}[1]{}
\newcommand{\sv}{$\clubsuit$}
\newcommand{\mbp}{$\spadesuit$}
\title{Many Body physics and the capacity of quantum channels with memory.}


\author{M. B. Plenio$^{1,2}$ \& S. Virmani$^{1,2,3}$}
\affiliation{$^{1}$ QOLS, Blackett Laboratory, Imperial College London,
Prince Consort Road, London SW7 2BW, UK}

\affiliation{$^{2}$ Institute for Mathematical Sciences, Imperial College
London, 53 Exhibition Road, London SW7 2PG, UK}

\affiliation{$^{3}$ STRI, University of Hertfordshire, College Lane, Hatfield, AL10 9AB}

\date{\today}

\begin{abstract}
In most studies of the capacity of quantum channels, it is
assumed that the errors in each use of the channel are
independent. However, recent work has begun to investigate the
effects of memory or correlations in the error, and has led to suggestions
that there can be interesting non-analytic behaviour in the capacity of
such channels. In a previous paper we pursued this issue by
connecting the study of channel capacities under correlated error to the study of critical
behaviour in many-body physics. This connection enables the use of
techniques from many-body physics to either completely solve or
understand qualitatively a number of interesting models of
correlated error with analogous behaviour to associated many-body systems.
However, in order for this approach to work rigorously, there are a number of technical properties
that need to be established for the lattice systems being considered.
In this article we discuss these properties in detail, and establish them
for some classes of many-body system.
\end{abstract}

\maketitle

\section{Introduction}

One of the most important problems of quantum information theory is
to try to determine the {\it channel capacity} of noisy quantum
channels. In a typical scenario, Alice would like to send Bob
information over many uses of a noisy quantum communication link. As
the channel is noisy, this cannot usually be done perfectly, and so
they must use some form of block encoding to combat errors. The
channel capacity is defined as the optimal rate at which information
may be transferred with vanishing error in the limit of a large
number of channel uses. There are a variety of different capacities,
depending upon whether Alice and Bob are interested in transmitting
classical or quantum information, and whether they have extra
resources such as prior entanglement. In this paper we will be
concerned mostly with the capacity for sending {\it quantum}
information, and so whenever we write the term `channel capacity' we
will implicitly be referring to the {\it quantum} channel capacity.

In most work on these problems, it has usually been assumed that the
noisy channel acts independently and identically for each channel
use. In this situation the transformation ${\cal E}_n$ corresponding
to $n$-uses of the channel may be written as an $n$-fold tensor
product of the single-use channel ${\cal E}_1$:
\begin{equation}
{\cal E}_n = {\cal E}_1 \otimes {\cal E}_1 \otimes ... \otimes {\cal
E}_1.
\end{equation}
However, in real physical situations there may be correlations in
the noise that acts between successive uses, an
interesting example being the decoherence of photons optical
fibres under the action of varying birefringence, which can be correlated due to mechanical
motion or slow temperature fluctuations \cite{Banaszek}. In such
situations one cannot describe the action of the channel in a simple
tensor product form:
\begin{equation}
{\cal E}_n \neq {\cal E}_1 \otimes {\cal E}_1 \otimes ... \otimes
{\cal E}_1.
\end{equation}
In this setting one must really describe the action of the channel
by a family of quantum operations corresponding to each number of
uses of the channel $n=1,2,...\infty$:
\begin{equation}
\{ {\cal E}_n \}_n, \label{family}
\end{equation}
We will call any such family of operations a {\it memory channel} or
a {\it correlated channel} \footnote{The term `{it correlated}' is
sometimes more appropriate as we will also discuss the notion of
correlated error in channels with a 2 or 3 spatial dimensional
structure, such as might arise in `egg box' storage such as optical
lattices. In such cases `memory' does not really have a meaning.}.
Defining the notion of channel capacity for such a correlated
channel is not always straightforward. In principle a family of
channels such as (\ref{family}) may not have any sensible limiting
behaviour as $n \rightarrow \infty$ \cite{caveat}. However, in this
paper we will not need to discuss this issue in detail, as we will
only consider fairly regular channels that have a (unique)
well-defined notion of channel capacity.

In the case of uncorrelated errors, it has recently been shown
\cite{Devetak} that the quantum channel capacity of an uncorrelated
quantum channel is given by:
\begin{equation}
Q({\cal E}) = \lim_{n \rightarrow \infty} {I({\cal E}^{\otimes n})
\over n} \label{qform}
\end{equation}
where $I(\xi)$ is the so-called {\it coherent information} of the
quantum channel $\xi$:
\begin{equation}
I(\xi) := \sup_{\rho} S(\xi(\rho)) - S(I \otimes \xi (|\psi \rangle
\langle \psi|))
\end{equation}
where $S$ denotes the von-Neumann entropy, $\rho$ is a state, and
$|\psi\rangle\langle \psi|$ is a purification of $\rho$.

Given that equation (\ref{qform}) is the quantum channel capacity
for memoryless channels, it is natural to hope that the
corresponding expression:
\begin{equation}
Q(\{{\cal E}_n\}) := \lim_{n \rightarrow \infty} {I({\cal E}_n)
\over n} \label{qcorr}
\end{equation}
will represent the quantum channel capacity in the case of
correlated errors. However, this will not always be the case, not least because
this limit does not always exist \cite{Kretschmann W,caveat}.
However, in this paper we will not only assume that this limit
exists, we will also initially work under the assumption that it
represents the true quantum channel capacity. We will later discuss
this assumption in some detail.

A similar situation occurs for the {\it classical} capacity of
correlated quantum channels, where formulae
(\ref{qform},\ref{qcorr}) can be replaced with similar expressions
involving the Holevo quantity instead of the coherent information.
Most prior work on calculating the capacities of correlated quantum
channels has focussed on the capacity for classical information.
Numerical and mathematical experiments involving a small number of
channel uses suggest that in a variety of interesting cases the
classical capacity of correlated channels can display interesting
non-analytic behaviour. For instance, the sequence of papers \cite{Macchiavello
P 02,Macchiavello PV 03, Daems} investigates a certain family of correlated
channels parameterized by a {\it memory factor} $\mu \in [0,1]$
which measures the degree of correlations. The results of
\cite{Macchiavello P 02,Macchiavello PV 03, Daems} demonstrate that when
the correlated channel is refreshed after every two uses (i.e.
consider ${\cal E}_2 \otimes {\cal E}_2 \otimes {\cal E}_2
\otimes...$, rather than the full correlated channel $\{{\cal
E}_n\}$), then there is a certain transition value $\mu=\mu_0$ at
which the channel capacity displays a definite kink, and above this
threshold the optimal encoding states suddenly change from product
to highly entangled. Similar phenomena have subsequently been
observed in a variety of other cases \cite{Karimipour,Karpov}.

Despite these interesting observations, it is still an open question
whether the sharp kinks in the capacity of these models still persist if the full
correlated channel $\{{\cal E}_n\}$ is considered as $n \rightarrow
\infty$, or whether this behaviour is just an artefact of the
truncation of the channel at low $n$. The main difficulty in
deciding such questions is that even under the assumption that
equations such as (\ref{qcorr}) (or its analogue for classical
information - the regularised Holevo bound) represent the true
quantum capacity of a given correlated channel $\{{\cal E}_n\}$, in
most cases such variational expressions are extremely difficult to
compute. It is however interesting to note that the non-analytic
behaviour observed in the channel capacity of correlated channels is
somewhat reminiscent of the non-analyticity of physical observables
that define a (quantum) phase-transition in strongly interacting
(quantum) many-body systems, where in contrast true {\it phase
transitions} usually {\it only} occur in the $n \rightarrow \infty$
limit.

Motivated by this heuristic similarity, in a previous paper \cite{Plenio V 07} we
connected the study of channels with memory to the study of many-body
physics. One advantage of this approach is that allows the construction of
 a variety of interesting examples of channels for which equation (\ref{qcorr}) can either be
understood qualitatively or even calculated exactly using the
techniques of many-body physics. One would otherwise usually expect regularized equations such as
(\ref{qcorr}) to either be quite trivial or completely intractable.
 This is perhaps the most important
consequence of this line of attack -- by relating correlated channels
directly to many-body physics, we obtain a good method for
displaying models of channels with memory that tread the interesting
line between `solvability' and `non-solvability', in analogy with
the many such statistical physics models that have been proposed
over the years. It is quite possible that the insights of {\it
universality}, {\it scaling}, and {\it renormalisation} that have
been so successful in many-body theory may provide valuable
intuition for the study of channels with correlated error.

Another advantage of this approach is its connection to
physically realistic models of correlated error. One can imagine
that in many real forms of quantum memory, such as optical lattices,
any correlated errors might originate from interaction with a
correlated environment and thus be strongly related to models of
statistical physics. This provides further physical motivation to
examine the properties of correlated channels with a many-body
flavour.

The connection to many body physics also naturally leads one
to consider channels with structure in 2 or more spatial
dimensions. In such situations it is no-longer appropriate to think
of correlations as `memory', as the correlations arise not through a
single time dimension, but perhaps through spatial proximity in more
than one dimension. In order to define a capacity in such
multidimensional situations one would have to decide how to quantify
the size of the channel. Natural options could include the total
number of particles in the system, or perhaps the size of one linear
dimension. Although we will not explicitly discuss multi-dimensional
examples in this work, such situations might have interesting connections
to the study of error tolerance in computational devices.

This paper is structured as follows. In order to make the paper
self-contained, in the sections preceding \ref{sec:conditions}
we present, including all missing detail, the results
of \cite{Plenio V 07}. In section \ref{sec:conditions} we discuss
in detail some sufficient conditions that many-body systems
must satisfy in order to lead to capacity results according to
the approach that we adopt - the arguments that lead to the
development of these conditions were sketched in \cite{Plenio V 07},
however here we provide the full argument. In sections \ref{proof1},
\ref{proof2} we prove that these conditions hold for finitely
correlated states and formulate a Fannes type inequality to
show the same result for harmonic chains. In the remaining
sections we discuss generalisations of our approach and
present conclusions.

\section{Many-body correlated channels.}

\begin{figure}[t]
\resizebox{8.5cm}{!}{\includegraphics{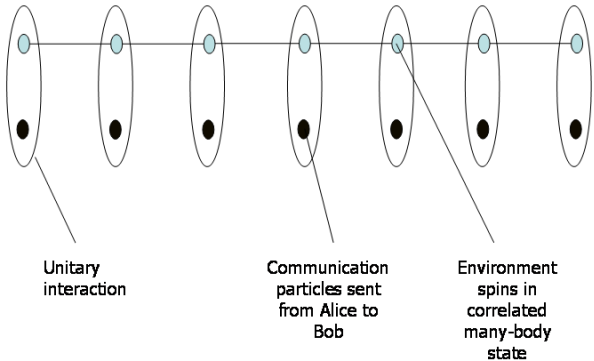}} \caption{
Each particle that Alice sends to Bob interacts with a separate
environmental particle from a many-body system.}
\label{modelpicture}
\end{figure}

In this section we re-cap the approach taken in \cite{Plenio V 07}
to construct correlated error models with links to many-body
physics. The starting point is to suppose, as usual, that
Alice transmits a sequence of particles to Bob (the `system'
particles), and that each particle interacts via a unitary $U$
with its {\it own} environmental particle. So far this is
exactly the same setting as uncorrelated noise. However, although
each system particle has its own separate environment, one can
introduce memory effects by asserting that the environment particles
are in the thermal/ground state of a many-body Hamiltonian, such
that the interaction terms lead to correlations in the environmental
state (see figure \ref{modelpicture}). Unlike the uncorrelated case,
this means that there will be correlations in the noise on different
system particles.
At this point it is important to discuss some of the subtleties
involved in the way that the `many-body' system was defined in
\cite{Plenio V 07}. In basic approaches to many-body physics, it
is usual to consider a system with a finite number of particles,
obtain thermal states and ground states, and then take a limit
as the number of particles is taken to infinity. In more mathematical
statistical physics literature
\cite{Bratelli R}, however, it is usual to consider genuinely
infinite systems from the start. This involves a number of technical
implications, including a very different approach to the concept of
a state, which can no longer be expressed in terms of basic density
matrices. The two approaches are not necessarily equivalent and may
lead to different results. To avoid such technicalities in this work
we will follow the former approach, and for each number of uses of
the channel $n$, we will consider a many-body system of size $n$. As
a family of channels for each $n$ this is a mathematically well
defined object, and it is a reasonable question to ask what the
resulting channel capacity is. In later sections of the paper we
will also assume {\it periodic} boundary conditions to enable us to
analyze whether equation (\ref{key}) is a valid quantum capacity or
not. Again, although this seems like an unnatural assertion, it is
mathematically well defined, and in many systems the boundary
conditions are believed to make a vanishingly small difference which
disappears in the large $n$ limit.

Of course even with these simplifications not all many-body systems
can be solved exactly, or even understood qualitatively. Moreover,
even if the many-body system can be well understood, the computation
of the limit (\ref{qcorr}) may still be difficult, and may depend
strongly upon the choice of the unitary $U$ describing the
interaction of each system particle with its associated
environmental particle. In order to provide concrete examples, one
must hence make a judicious choice of $U$ in order to make
analytical progress. As in \cite{Plenio V 07}, we choose $U$ to be of the form of a
controlled-unitary interaction, where the environmental particles
act as controls.
In fact, for ease of explanation we will also
initially restrict the system and environment particles to be
2-level spins, and the interaction $U$ to be a controlled-phase
(`CPHASE') gate, which in the computational basis for 2-qubits is
defined as,
\begin{eqnarray}{\rm CPHASE} = \left(\begin{array}{cccc}
     1 & 0 & 0 & 0 \\
     0 & 1 & 0 & 0 \\
     0 & 0 & 1 & 0 \\
     0 & 0 & 0 & -1 \end{array}\right).\end{eqnarray}
Later we will discuss how higher level analogues of the CPHASE
enable similar connections to many-body theories with
constituent particles with a higher number of levels.
The reason we make these choices  for the controlled unitary interactions is that explicit formulae
may be derived for the capacity in terms of relatively simple {\it entropic} expressions
which are especially amenable to analysis.

The restriction to controlled-unitary interactions also enables us
to consider environment particles that are classical. For instance,
in the case of classical environment 2-level spins, the `CPHASE'
interaction will be taken to mean that the system qubit undergoes a
Pauli-$Z$ rotation when the environment spin is up, otherwise it is
left alone. It turns out that by considering classical
environments it is possible to make more direct connections between
the channel capacity of our models and concepts from statistical
physics.

So let us proceed in trying to understand the capacity in cases in
which the system particles are all two-level systems, with a CPHASE
interaction. It is helpful to write the resulting channels in a more
explicit form. Let us consider a quantum environment first. Let
$|0\rangle$ denote spin-down, and $|1\rangle$ denote spin-up. Let us
suppose that the environment consists of $N$ spins (eventually we
will be interested in the limit $N \rightarrow \infty$) initially in
a state:
\begin{equation}
\sum_{\bf{x},\bf{y}} \rho_{\bf{x},\bf{y}} |x_1 x_2 ... x_N \rangle
\langle y_1 y_2 ... y_N | \label{envstate}
\end{equation}
where the sum is taken over all N-bit strings ${\bf x},{\bf y}$, and
$x_j/y_j$ denote the $j$th bit of strings ${\bf x}/{\bf y}$
respectively. We can also describe a classical environment in the
same way, simply by restricting the input environment state $\rho$
to be diagonal in the computational basis - the CPHASE interaction
will in this case leave the environment unchanged, and will affect
the system qubits as if the controls are entirely classical.

If the environment is in the state (\ref{envstate}), and the system
qubits are initially in the state $\sigma$, then the channel acting
upon the system qubits is given by:
\begin{equation}
\sigma \rightarrow \sum_{\bf{x}} \rho_{\bf{x},\bf{x}} Z_1^{x_1}
Z_2^{x_2} ... Z_N^{x_N} \sigma (Z_1^{x_1} Z_2^{x_2} ...
Z_N^{x_N})^\dag \label{channel}
\end{equation}
where $Z_i$ denotes the Pauli-Z operator acting upon qubit $i$.
Hence, regardless of whether the environment is considered quantum
or classical, the channel that we have described is a probabilistic
application of $Z$-rotations on various qubits. Although we will
consider qubit $i$ to be transmitted earlier in time than any other
qubit $j$ with $i < j$, there is no need for us to actually impose
such a time ordering - because all the CPHASE interactions commute
with each other, such time ordering is irrelevant \cite{causal}.

We will be interested in computing equation (\ref{qcorr}) for such
many-body correlated channels. In the next section we will show that
the channel capacity of this channel is given by a simple function
of the entropy of the {\it diagonal} elements in the spin up/down
basis of the environmental state, i.e.
\begin{equation}
- \sum_{{\bf x}} \rho_{{\bf x},{\bf x}}\log \rho_{{\bf x},{\bf x}}.
\end{equation}
In the case of a classical environment this is just the actual
entropy of the spin-chain. This observation is very useful, as it
allows us to apply all the formalism of many-body physics to the
problem, also enabling us to that intuition to observe a number of
interesting effects. In the quantum case this function does not
correspond to a conventional thermodynamic property, however, we
will discuss examples where it is still amenable to a great deal of
analysis using many-body methods.

\section{A formula for the coherent information of our models.}

In order to calculate the regularised coherent information
(\ref{qcorr}) for our many-body correlated channels, we will utilise
the close relationship between the quantum channel capacity and the
entanglement measure known as the {\it Distillable Entanglement}
\cite{distill}. This connection utilises a well known mapping
between quantum operations and quantum states. Given any quantum
operation ${\cal{E}}$ acting upon a $d$-level quantum system, one
may form the quantum state:
\begin{equation}
J({\cal{E}}) = I \otimes {\cal{E}}(|+\rangle\langle+|) \label{cj}
\end{equation}
where $|+\rangle = {1 \over \sqrt{d}} \sum_{i=1..d}|ii\rangle$ is
the canonical maximally entangled state of two d-level systems. The
state $J({\cal{E}})$ is sometimes referred to as the {\it
Choi-Jamiolkowski} state of the operation ${\cal{E}}$ \cite{CJ}. It can be
shown that the mapping from ${\cal{E}}$ to $J({\cal{E}})$ is
invertible, and hence the state $J({\cal{E}})$ gives a one-to-one
representation of a quantum operation. We will show that for the
kinds of correlated error channel that we have described above in
equation (\ref{channel}), the quantum channel capacity
$Q({\cal{E}})$ of the channel equals $D(J({\cal{E}}))$, the
distillable entanglement of the state $J({\cal{E}})$.

To make the presentation more transparent, we will make the argument
for the Choi-Jamiolkowski (CJ) state of a particular single qubit
channel, as it is straightforward to generalize the argument to the
entire family of memory-channels described above. Hence let us
consider the following single qubit `dephasing' channel:
\begin{equation}
{\cal{E}}: \rho \rightarrow p \rho + (1-p) Z \rho Z^\dag \label{cov}
\end{equation}
where $p$ is a probability, and $Z$ is the Pauli Z operator. The CJ
representation of this channel is:
\begin{equation}
J({\cal{E}}) = I \otimes {\cal{E}}(|+\rangle\langle+|)
\end{equation}
where $|+\rangle$ is chosen as in equation (\ref{cj}).

The argument relies upon the fact that the channel ($\ref{cov}$)
possesses some useful symmetry. This symmetry leads to the property
that having one use of the channel is both mathematically and {\it
physically} equivalent to having one copy of $J({\cal{E}})$. Suppose
that you have one use of ${\cal{E}}$, you can easily create
$J({\cal{E}})$. However, it turns out that with one copy
$J({\cal{E}})$ you can also implement one use of ${\cal{E}}$. Hence
both the operation and the CJ state are physically equivalent
resources. The argument works as follows. Suppose that you have
$J({\cal{E}})$ and you want to implement one action of ${\cal{E}}$
upon an input state $\rho$. This can be achieved by teleporting
$\rho$ through your copy of $J({\cal{E}})$. This will leave you with
the state ${\cal{E}}(\sigma_i \rho \sigma^\dag_i )$, with the Pauli
operator $\sigma_i$ depending upon the outcome of the Bell
measurement that does the teleportation. However, the channel
($\ref{cov}$) commutes with all Pauli rotations. So we can ``undo"
the effect of the Pauli by applying the inverse of $\sigma_i$, which
for Paulis is just $\sigma_i$ itself. Hence we have: $\sigma_i
{\cal{E}}(\sigma_i \rho \sigma^\dag_i ) \sigma_i = {\cal{E}}(\rho)$.
Hence by teleporting into $J({\cal{E}})$ and undoing the Pauli at
the end we can implement one use of the operation.

This observation allows us to relate the channel capacity of the
channel to the distillable entanglement of the CJ state. The proof
proceeds in two steps, and follows well known ideas taken from
articles such as \cite{distill}. The aim is to show that the 1-way
distillable entanglement of $J({\cal E})$ is equivalent to $Q({\cal
E})$, so that previous results on $D(J({\cal E}))$ may be applied.

\medskip

\smallskip (1) PROOF THAT $Q({\cal{E}})$ $\leq$ 1-way distillation: (1) Alice
prepares many perfect EPR pairs and encodes one half according to
the code that achieves the quantum capacity $Q({\cal{E}})$. (2) She
teleports the encoded qubits through the copies of $J({\cal{E}})$,
telling Bob the outcome so that he can undo the effect of the
Paulis. (3) This effectively transports all encoded qubits to Bob,
at the same time acting on them with ${\cal{E}}$. (4) Bob does the
decoding of the optimal code, thereby sharing perfect EPR pairs with
Alice, at the rate determined by $Q({\cal{E}})$. As this is a
specific one-way distillation protocol, this means that $Q \leq D$.

\medskip

\smallskip (2) PROOF THAT $Q({\cal{E}})$ $\geq$ 1-way distillation: (1) Alice
prepares many perfect EPR pairs and sends one half of each pair
through many uses of the channel ${\cal{E}}$, (2) She and Bob do one
way distillation of the resulting pairs (this involves only forward
classical communication from Alice to Bob). (3) Thereby they share
perfect EPR pairs, at the rate determined by $D(J({\cal{E}}))$, the
1-way distillable entanglement of . (4) They can use these EPR pairs
to teleport qubits from Alice to Bob. As this is a specific quantum
communication protocol, this means that $Q \geq D$.

\medskip

These arguments can easily be extended to apply to any channel that
is a mixture of Pauli rotations on many qubits, hence including the
memory channel models that we have described above. Hence to
calculate the quantum channel capacity of our channels we must
calculate the Distillable entanglement of the channel's CJ state.
Fortunately, the CJ state of our channel is a so-called {\it
maximally correlated} state, for which the distillable entanglement
is known to be equivalent to the Hashing bound:
\begin{equation}
D(J({\cal E})) = S({\rm tr_B}\{J({\cal E})\}) - S(J({\cal E}))
\end{equation}
where $S$ is the von-Neumann entropy. Note that for such channels
${\cal E}$ this expression is equivalent to the {\it single copy}
coherent information, which is hence additive for product channels
${\cal E}^{\otimes n}$. In our case we are interested in the
regularised value of this quantity for correlated channels, i.e.:
\begin{equation}
Q(\{{\cal E}_n\}) = \lim_{n \rightarrow \infty} {D(J({\cal E}))
\over n}= \lim_{n \rightarrow \infty} {S(J({\cal E}_n)_A) -
S(J({\cal E}_n)) \over n}
\end{equation}
which can be computed quite easily as:
\begin{equation}
Q(\{{\cal E}_n\}) = 1 - \lim_{n \rightarrow \infty} {S({\rm
Diag}(\rho_{env})) \over n} \label{key}
\end{equation}
where ${\rm Diag}(\rho_{env})$ the state obtained by eliminating all
off-diagonal elements of the state of the environment (in the
computational basis). Hence the computation of the quantum channel
capacity of our channel $\{{\cal E}_n\}$ reduces to the computation
of the regularised diagonal entropy in the limit of an infinite spin
chain. Although in most cases this quantity is unlikely to be
computable analytically, it is amenable to a great deal of analysis
using the techniques of many-body theory. It is also interesting to note
the intuitive connection between expression (\ref{key}) and work
on environment assisted capacities - in the case of random unitary
channels, where the unitaries are mutually orthogonal, the diagonal entropy in
expression (\ref{key}) has a natural interpretation as the amount of classical
information that needs to be recovered from the environment in order
to correct the errors \cite{Gregoratti,Buscemi}.

Although the above analysis has been conducted for 2-level
particles, it can be extended to situations involving $d$-level
systems. In the $d$-level case one can replace CPHASE with a
controlled shift operation of the form:
\begin{equation}
\sum_{i=1..d} |k \rangle\langle k| \otimes Z(k)
\end{equation}
where the $Z(k)= \sum_j \exp(i2\pi kj/d) |j \rangle\langle j|$ are
the versions of the qubit phase gate generalized to $d$-level
systems, and the first part of the tensor product acts on the
environment. With this interaction all the previous analysis goes
through, and the $d$-level version of eq. (\ref{key}):
\begin{equation}
Q(\{{\cal E}_n\}) = \log (d) - \lim_{n \rightarrow \infty} {S({\rm
Diag}(\rho_{env})) \over n} \label{dkey}
\end{equation}
gives the regularized coherent information, where Diag($\rho$)
refers to the diagonal elements in the $d$-level computational
basis. It is important to consider the generalization to $d$-level
systems because the thermodynamic properties of many-body systems do
not always extend straightforwardly to systems with a higher number
of levels. For instance, one possible generalization of the Ising
model to $d$-level systems is the {\it Potts} model, which leads to
some very interesting and non-trivial mathematical structure
\cite{Wu review}, and in the quantum Heisenberg model the presence
of a ground state gap depends on where the spins in the chain are
integral or half-integral \cite{Sachdev}.

The simplicity of equation (\ref{key}) enables one to immediately
write down many noise models for which the regularized coherent
information can both be calculated, and also represents the quantum
channel capacity of the correlated channel. In particular, let us
suppose that the environment consists of classical systems described
by a classical {\it Markov Chain} (those readers not familiar with
the Markov chain terminology required here please see chapter 5 of
\cite{Welsh} for a very readable introduction). If the state at each
`site' $s$ in the environment represents the instantaneous state of
a Markov chain at time $s$, then the regularised entropy in equation
(\ref{dkey}) is given by the {\it entropy rate} of the Markov chain
\cite{Welsh}, provided that the Markov process is both {\it
irreducible} \footnote{{\it Irreducibility} means that given any
starting state there is a non-zero probability of eventually going
through any other state.} and possesses a {\it unique} stationary
(equilibrium) state. Let the transition matrix of $M$ of the Markov
chain be defined such that $p_i(s+1)= \sum_j M_{ij}p(s)_j$, let
$v_i$ be the $i$th element of the stationary probability
distribution, and let $H_i$ be the entropy of column $i$ in the
Markov chain transition matrix. With these conventions the entropy
rate is given by:
\begin{equation}
\lim_{n \rightarrow \infty} {S({\rm Diag}(\rho_{env})) \over n} =
\sum_{i=1,..,d} v_i H_i
\end{equation}
In these cases the correlated channels fit quite neatly into the
class of models proposed in \cite{Bowen M,Kretschmann W}, and
moreover these channels will be {\it forgetful} \cite{Kretschmann
W}. As proven in \cite{Kretschmann W}, for forgetful channels the
regularized coherent information is equal to the quantum capacity
(see \cite{Hamada} for an independent coding argument which also works
for Markov chain channels implementing generalised Pauli rotations).
Hence for these models equation (\ref{dkey}) represents the true
quantum channel capacity, and so we may write explicitly:
\begin{equation}
Q({\rm Markov}) = \log(d) - \sum_{i=1,..,d} v_i H_i. \label{markov}
\end{equation}
When unique, the stationary distribution of a Markov chain is given
by the unique maximal right eigenvector (of eigenvalue 1) of the
transition matrix. Related results have been obtained independently in
the articles \cite{Hamada,Darrigo}.


\section{Environment that is a classical system}

In the case of a classical environment, the second term of equation
(\ref{key}) is precisely the entropy of the environment, and so it
can easily be computed in terms of the partition function.

The partition function of the classical system is defined as:
\begin{equation}
Z = \sum_i \exp (- \beta E_i)
\end{equation}
where the $E_i$ are the energies of the various possible
configurations, and $\beta = 1/(k_B T)$, with $T$ the temperature
and $k_B$ Boltzmann's constant. The entropy (in nats) of the system
is given by the following expression:
\begin{eqnarray}
S({\rm Diag}(\rho_{env})) = \left( 1 - \beta {\partial \over
\partial \beta} \right) \ln Z.
\end{eqnarray}
This means that in the case of a classical environment our channel
capacity becomes
\begin{equation}
Q(\{{\cal E}_n\}) = 1 -  \log_2(e) \left( 1 - \beta {\partial \over
\partial \beta} \right) \lim_{n \rightarrow \infty} {1 \over n} \ln
Z
\end{equation}
where the $ \log_2(e)$ converts us back from nats to bits. This
expression means that we can use all the machinery from classical
statistical mechanics to compute the channel capacity.

In particular, any spin-chain models from classical physics that can
be solved exactly will lead to channels with memory that can be
`solved exactly' (provided that one can show that the regularized
coherent information is indeed the capacity, a problem that we shall
discuss in later sections). The most famous example of an `exactly
solvable' classical spin-chain model is the Ising model. We will
discuss the classical Ising model in detail in the next section, as
it will also be relevant to a certain class of quantum spin-chains.

However, there are also {\it many} classical spin-chain models that
{\it cannot} always be solved exactly, but which can be connected to
a wide variety of physically relevant models with interesting
behaviour. As just one example, consider modifying the Ising
spin-chain model to allow exponentially decaying interactions
between non-adjacent spins. The resulting model can be related to a
quantum double-well system, and is also known to exhibit a phase
transition \cite{expising}. This means that the corresponding
correlated channels will also exhibit similar behaviour, provided of
course that the limit eq. (\ref{qcorr}) truly represents the quantum
channel capacity for the models.

In this paper we will not give detailed discussion of any further
models involving a classical environment (other than the classical
1D Ising chain, which we will discuss in the next section). As our
expression (\ref{key}) is simply the entropy of the classical
environment, the interested reader may simply refer to the many
interesting classical models (both solvable and almost solvable)
that are well documented in the literature. Of course, to make the
analysis rigorous one would need to show that expression
(\ref{qcorr}) is the formula for the quantum capacity in these
cases. However, we conjecture that for most sensible models this
should be true. In the final section of the paper we will present an
analysis that demonstrates this for a family of 1D models.

\section{Quantum Environments}

Unfortunately expression (\ref{key}) does not correspond to a
standard thermodynamic function of the environment state when the
environment is modelled as a quantum system. It represents the
entropy of the state that results when the environment is decohered
by a dephasing operation on every qubit. Although this quantity is
not typically considered by condensed matter physicists, there is
some hope that it will be amenable to analysis using the techniques
of many-body theory.

In this paper we will make a small step towards justifying this hope
by analytically considering a class of quantum environments inspired
by recent work on so-called {\it Finitely Correlated} or {\it Matrix
Product States} \cite{Fannes NW 92}.

\begin{figure}[t]
\resizebox{8.5cm}{!}{\includegraphics{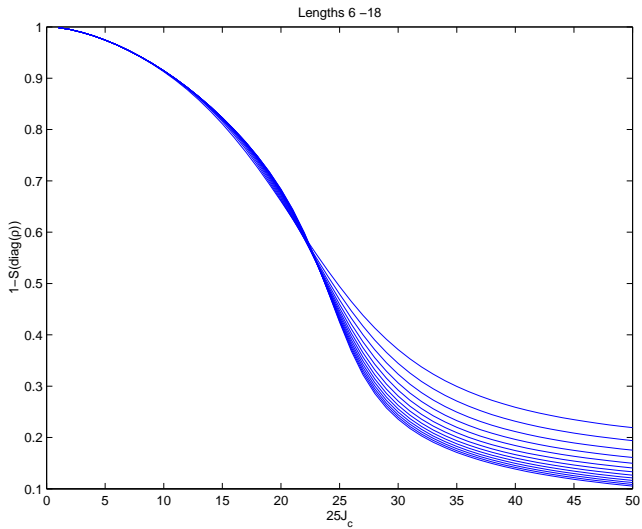}}
\caption{Numerics for the quantum Ising model suggest that there may
be transition behaviour in the capacity at the phase transition
point of the 1D quantum Ising model. In this figure the central
point of the horizontal axis is the transition point of the quantum
Ising model, and the curves become increasingly steep as the number
of spins is increased from 6 to 18. As the quantum Ising model can
be solved exactly in 1D, it is quite possible that an analytical
solution may be found for the channel capacity.}
\label{quantumising}
\end{figure}

We will leave attempts to analytically study more complicated models
to another occasion, although in figure (\ref{quantumising}) we
present some numerical evidence that the quantum 1D Ising model
displays a sharp change in capacity at the transition point.

\section{Quantum capacity for finitely correlated environments described by rank-1
matrices}\label{example}

{\it Finitely Correlated} or {\it Matrix Product} states are a
special class of efficiently describable quantum states that have
provided many useful insights into the nature of complex quantum
systems \cite{Fannes NW 92}. In a recent paper \cite{Wolf OVC
05} it has been demonstrated that a variety of interesting
Hamiltonians can be constructed with exact matrix product ground
states, such that the Hamiltonians in question undergo non-standard
forms of quantum `phase transition'.

As matrix product states are relatively simple to describe, one
might hope that for such ground states the computation of equation
(\ref{key}) may be particularly tractable. In this section we will
see that for matrix product states involving {\it rank-1} matrices
the analysis is particularly simple, and may be reduced to the
solution of a classical 1D Ising model.

Let us consider a 1D matrix product state, where each particle is a
2-level quantum system, $|0\rangle,|1\rangle$. Let us assume that
the matrices associated to each level are independent of the site
label, and are given by $Q_0$ for level $|0\rangle$ and $Q_1$ for
level $|1\rangle$. Hence the total {\it unnormalised} state can be
written as:
\begin{equation}
|\psi\rangle = \sum_{i,j,k.. \in \{0,1\}} {\rm{tr}}\{Q_iQ_jQ_k...\}
|ijk...\rangle
\end{equation}
From the form of expression (\ref{key}) we see that we are only
interested in the weights of the diagonal elements in the
computational basis, or equivalently the state that results from
dephasing each qubit. It is easy to see that this {\it unnormalised}
state will be given by:
\begin{equation}
\rho = \sum_{i,j,.. \in \{0,1\}} {\rm{tr}}\{(Q_i \otimes Q^*_i)(Q_j
\otimes Q^*_j)...\} |ij...\rangle\langle ij...|.
\end{equation}
In this expression if we relabel the matrices $A=Q_0 \otimes Q^*_0$
and $B=Q_1 \otimes Q^*_1$ then probability of getting various
outcomes when measuring the environment in the computational basis
will be given by traces of all possible products of the $A$s and
$B$s. For instance, the probability of getting 01100... when
measuring the environment in the computational basis will be given
by:
\begin{equation}
p_{01100...} ={1 \over C(N)} \mbox{tr} \{ ABBAA... \}
\end{equation}
where $N$ is the number of qubits in the environment, and $C(N)$ is
a normalisation factor given by:
\begin{equation}
C(N) = \mbox{tr} \{ (A+B)^N \}.
\end{equation}
$C(N)$ can be computed by diagonalisation. In the rest of this
section we will be interested in cases where $A$ and $B$ are both
square {\it rank-1} matrices. Some of the example Hamiltonians
discussed in \cite{Wolf OVC 05} have ground states with this
property, and in fact some special cases of the noise models
presented in \cite{Macchiavello P 02,Macchiavello PV
03,Karimipour,Karpov} can also be expressed in the form of matrix
product environments with rank-1 matrices (although in general those
models require more than two matrices as they require environmental
spins with more than 2 levels). We will show that in such situations
the diagonal entropy in the computational basis is equivalent to the
entropy of a related classical Ising chain.

\begin{figure}[t]
\resizebox{8.5cm}{!}{\includegraphics{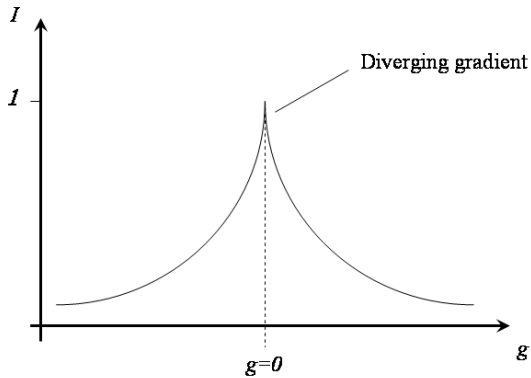}} \caption{This
schematic figure shows the channel capacity when the environment is
the ground state of the Hamiltonian given in equation
(\ref{wolfham}). The symmetry in this plot is to be expected as the
channel is invariant under the replacement $g \rightarrow -g$.
However, near the `phase transition' point $g=0$, the gradient
diverges.} \label{wolfplot}
\end{figure}

The first thing to note is that rank-1 matrices are almost
idempotent. In fact, if $A,B$ are both rank-1 matrices, then we have
that:
\begin{equation}
A^n = a^{n-1}A ~~~~~~~;~~~~~~~ B^m = b^{m-1}B
\end{equation}
where $a$ is the only non-zero eigenvalue of $A$, and $b$ is the
only non-zero eigenvalue of $B$. Note that because of the form of $A$ and
$B$ as the tensor product of a matrix and its complex conjugate, these eigenvalues $a,b$ must be non-negative. We can define the
normalised matrices:
\begin{equation}
\tilde{A} = {A \over a}~~~~~~~;~~~~~~~ \tilde{B} = {B \over b}.
\end{equation}
These normalised matrices {\it are} idempotent. To see how this can
help, consider a particular string, say,
\begin{eqnarray}
p_{0111000} = {1 \over C(N)} \mbox{tr} \left\{\left( ABBBAAA \right)
\right\} \nonumber
\end{eqnarray}
if we substitute $\tilde{A}$ and $\tilde{B}$ into this expression,
and use the idempotency, then the strings of consecutive As and Bs
will collapse to just one $\tilde{A}$ or $\tilde{B}$, with total
factors of $a^4$ and $b^{3}$ inserted outside the trace:
\begin{eqnarray}
p_{0111000} = & {1 \over C(N)} a^4 b^3\mbox{tr} \left\{\left(
\tilde{A}\tilde{B}\tilde{A} \right) \right\} \nonumber \\
& = {1 \over C(N)} a^4 b^3 \mbox{tr} \left\{ \left(
\tilde{A}\tilde{B} \right) \right\}
\end{eqnarray}
It is easy to see that this form is quite general - the probability
of getting a particular string will collapse to a simple expression.
If there are $l$ occurrences of $A$ and $n-l$ occurrences of $B$ in
the string, and $K$ counts the number of boundaries between blocks
of $A$s and blocks od $B$s, then the probability of the string
becomes:
\begin{eqnarray}
{1 \over C(N)} (a^{l}b^{N-l}) \mbox{tr} \left\{\left( {\tilde{A}
\tilde{B}}\right)^{K} \right\}. \nonumber
\end{eqnarray}
Noting that $\tilde{A}\tilde{B}$ will also be a rank-1 matrix, let
us use the letter $c$ to refer to its only non-zero eigenvalue.
Hence the probability becomes:
\begin{eqnarray}
 {1 \over C(N)} a^{l}b^{N-l} c^K \label{probs} 
\end{eqnarray}
This expression tells us quite a lot - firstly for any given channel
described by rank-1 MPS states, the only parameters
that matter are $a,b,c$. 
So we needn't work with the actual matrices defining our state, we
only need to work with matrices of our choosing that have the same
parameters $a,b$ and $c$. In the following we will assert
that $c$ is non-negative - this is guaranteed because of
the following argument: it holds that $c={\rm{tr}}\{\tilde{A}\tilde{B}\}$,
because $\tilde{A}\tilde{B}$ is rank-1, but because $\tilde{A}\tilde{B} = Q_0Q_1 \otimes Q^*_0Q^*_1/(ab)$,
where $a,b$ are non-negative, this means that $c$ must be non-negative.
So let us just go ahead
and pick the following matrices:
\begin{eqnarray}
  A = \left(\begin{array}{cc}
     a & \sqrt{c a b} \\
     0 & 0 \end{array}\right); B = \left(\begin{array}{cc}
     0 & 0 \\
     \sqrt{c a b} & b \end{array}\right) \label{choice}.
\end{eqnarray}
These matrices clearly have non-zero eigenvalues $a,b$ respectively.
So what about the eigenvalue of $\tilde{A}\tilde{B}$? For the above
choice of matrices we find that:
\begin{eqnarray}
  \tilde{A}\tilde{B} = \left(\begin{array}{cc}
     c & {\sqrt{c b \over a}} \\
     0 & 0 \end{array}\right).
\end{eqnarray}
Hence we find that the matrices that we have chosen have the correct
values of $a,b,c$, as required. Now we notice that the matrices that
we have chosen in equation (\ref{choice}) are very similar to the
matrices that would define a classical Ising chain. In fact, if we
make the following change of variables from $a,b,c$ to $J,D,M$:
\begin{eqnarray}
a = \exp(\beta(J+M)), \nonumber \\
b = \exp(\beta(J-M)), \nonumber \\
c = \exp(-\beta(4J+2D)).
\end{eqnarray}
The inverse transformations are:
\begin{eqnarray}
\beta J = (\ln (a) + \ln (b))/2, \nonumber \\
\beta M = (\ln (a) - \ln (b))/2, \nonumber \\
\beta D = -(\ln (a) + \ln (b)) - (1/2)\ln (c). \label{inverse}
\end{eqnarray}
It turns out that the parameters $J,D$ will represent coupling
constants and $M$ will represent a magnetic field. To see this let
us insert the new parameters into the choice of $A,B$ in equation
(\ref{choice}). Then we get that the matrices (\ref{choice}) can be
written:
\begin{eqnarray}
  A = \left(\begin{array}{cc}
     \exp (\beta(J+M)) & \exp(-\beta(J+D)) \\
     0 & 0 \end{array}\right), \nonumber \\
     B = \left(\begin{array}{cc}
     0 & 0 \\
     \exp(-\beta(J+D)) & \exp(\beta(J-M)) \end{array}\right)
\end{eqnarray}
The matrices in such a rank-1 MPS are essentially the top row and
bottom row of a transfer matrix. Comparing these matrices to the
classical Ising transfer matrix we see that the following
Hamiltonian (where for convenience we now follow the usual physics
convention that $s_i \in \{-1,+1\}$):
\begin{eqnarray}
H &=& - \sum_i J s_i s_{i+1} - M s_i + D(1-s_i s_{i+1}) \nonumber \\
&=& - \sum_i (J-D) s_i s_{i+1} - M s_i + D
\end{eqnarray}
The $D$ is just a constant shift in spectrum, so we can simply
consider the Ising chain with Hamiltonian:
\begin{eqnarray}
H = - \sum_i (J-D) s_i s_{i+1} - M s_i
\end{eqnarray}
The partition function for such a chain of N particles depends upon
the transfer matrix for this (rescaled!) Hamiltonian:
\begin{eqnarray}
T = \left(\begin{array}{cc}
     \exp(\beta(J-D+M)) & \exp(-\beta(J-D)) \\
     \exp(-\beta(J-D)) & \exp(\beta(J-D-M)) \end{array}\right).
     \label{transfer}
\end{eqnarray}
Now from the partition function we can calculate the entropy, and
hence the capacity of our channel. The formula turns out to be:
\begin{eqnarray}
C &=& 1 -  \log_2(e) \left( 1 - \beta {\partial \over \partial
\beta}
\right) \lim_{N \rightarrow \infty} {1 \over N} \ln Z \nonumber\\
&=& 1 -  \log_2(e) \left( 1 - \beta {\partial \over \partial \beta}
\right) \ln \lambda_1
\end{eqnarray}
where $\lambda_1$ is the maximal eigenvalue of the transfer matrix
(\ref{transfer}). Using these equations and equation (\ref{inverse})
one can perform the (tedious) manipulation required to derive a
formula for the regularized coherent information in terms of the
coefficients $a,b,c$. Although we do not present the formula that is
obtained, figure (\ref{wolfplot}) shows the result for the model
Hamiltonian presented in \cite{Wolf OVC 05}:
\begin{eqnarray}
    H &=& \sum_i 2(g^2-1)\sigma_z^{(i)}\sigma_z^{(i+1)} -
    (1+g)^2\sigma_x^{(i)}\nonumber\\
    && + (g-1)^2\sigma_z^{(i-1)}\sigma_x^{(i)}\sigma_z^{(i+1)}\, .
\label{wolfham}
\end{eqnarray}
for which the ground state is known to be a matrix product state of
the form:
\begin{eqnarray}
    Q_0 &=& \left(\begin{array}{cc}   0 & 0\\
                                      1 & 1
                                      \end{array}\right)\;\;\;
    Q_1 = \left(\begin{array}{cc}   1 & g \\
                                      0 & 0
                                   \end{array}\right).
    \nonumber
\end{eqnarray}
This model system has a non-standard `phase transition' at $g=0$, at
which some correlation functions are continuous but
non-differentiable, while the ground state energy is actually
analytic \cite{Wolf OVC 05}. As discussed in the caption of figure
(\ref{wolfplot}), this behaviour is mirrored in the channel
capacity.

\section{Conditions under which the regularised coherent information represents the true capacity.}\label{sec:conditions}

In this section we will explore under what conditions our assumption
that the regularized coherent information of equation (\ref{qcorr}):
\begin{equation}
Q(\{{\cal E}_n\}) := \lim_{n \rightarrow \infty} {I({\cal E}_n)
\over n} \label{qrep}
\end{equation}
correctly represents the true quantum capacity of our correlated
channels, assuming of course that this limit exists. In the course
of the discussion we will also need to consider under what
conditions the regularized Holevo bound:
\begin{equation}
C({\cal E}_n):= \lim_{n \rightarrow \infty} {\chi({\cal E}_n) \over
n} \label{crep}
\end{equation}
represents the capacity of the channel for {\it classical}
information. The Holevo bound $\chi({\cal E})$ for a quantum channel
${\cal E}$ is defined as \cite{Nielsen C}:
\begin{equation}
\chi({\cal E}) = \sup_{\{p_i,\rho_i\}}  S({\cal E}(\sum_i p_i \rho_i)) -
\sum_i p_i S({\cal E}(\rho_i))
\end{equation}
where the supremum is taken over all probabilistic ensembles of
states $\{p_i,\rho_i\}$, and $S$ as usual represents the von Neumann
entropy. As pointed out in \cite{Bowen DM,Kretschmann W}, showing
that equations (\ref{qrep})/(\ref{crep}) are {\it upper} bounds to
the quantum/classical capacity of a correlated channel is
straightforward - one can use exactly the same arguments used in the
memoryless case \cite{Holevo,Schumacher W,Barnum NS,Devetak}.
Showing that equations (\ref{qrep})/(\ref{crep}) also give {\it
lower} bounds to the relevant capacities is not as simple, and may
not be true for some many-body environments.

However, it turns out that if the correlations in the many-body
system fall off sufficiently strongly, then the channel will be
reasonably well behaved and equation (\ref{qrep}) is true. In this
section we will make this statement quantitative. We will closely
follow the approach taken in \cite{Kretschmann W} in the analysis of
so-called {\it forgetful} channels. Some of the subtleties involved
in the analysis are explained in more detail in Section VI of that
paper. The conditions that we obtain are independent of the unitary which
governs the interaction between each system particle and its corresponding
environment, and so are applicable more widely than the dephasing interaction
considered here.

\subsection{A qualitative description of the argument.}

\begin{figure}[t]
\resizebox{8.5cm}{!}{\includegraphics{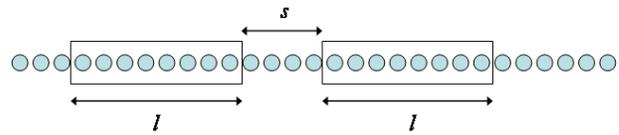}} \caption{
The Live blocks of length $l$ are separated by Spacer blocks of
length $s$. By discarding the spacer particles the channel
effectively becomes a product channel on the Live blocks.}
\label{livespace}
\end{figure}

In this subsection we present an intuitive sketch of the argument that we
will follow. Imagine that the
correlated channel is partitioned into large blocks that we shall
call {\it live} qubits, separated by small blocks that we shall call
{\it spacer} qubits. The idea is to throw away the spacer qubits,
inserting into them only some standard state, and to only use the
{\it live} qubits to encode information (see figure
\ref{livespace}). If we are to follow this procedure, then we will
not be interested in the full channel, but only in its effect upon
the live qubits. Let us use the phrase {\it live channel} to
describe the resulting channel - i.e. the reduced channel that acts
on the {\it live} qubits only. If the correlations in the many-body
system decay sufficiently strongly, then by throwing away just a few
spacer qubits we will find that the live channel closely
approximates (in a sense to be discussed later) a memoryless
channel. Let us call this memoryless channel the {\it product}
channel. One can imagine trying to use the codes that achieve the
capacity of the {\it product} channel, without any further
modifications, as codes for the {\it live} channel. It turns out
that under the `right conditions' these codes are not only good
codes for the live channel, but their achievable rates approach
equation (\ref{qrep}). The goal of the next subsection will be to explore
exactly what these `right conditions' are.

The quantitative arguments follow the method used in \cite{Kretschmann W},
where three steps are required to show that equation (\ref{qrep}) is an achievable
rate:
\begin{itemize}
\item[[A]] First we must show that product codes for the transmission of classical information
are good codes for the Live channel.
\item[[B]] Then we must show that these good codes allow the regularized Holevo quantity
to be an achievable rate. This is done by showing that the product channel Holevo
quantity (which can be achieved by product codes) essentially converges to the regularized
Holevo quantity for the whole channel.
\item[[C]] Then we must argue that these arguments for the transmission of classical
information can be `{\it coherentified}' (in the manner of \cite{Devetak}) to a good quantum code attaining equation (\ref{qrep}).
\end{itemize}
In the next subsection we go through this process in detail to derive sufficient conditions to
demonstrate the validity of equation (\ref{qrep}) for our many-body channels.

\subsection{Derivation of the conditions.}

In this subsection we will go through steps [A],[B],[C] in turn.

\subsubsection{Step [A]}

We will assume that the many-body systems in question satisfy
periodic boundary conditions and are translationally invariant (this
means that the corresponding correlated channel $\{{\cal{E}}_n\}$
does not quite fit into the definition of causality proposed by
\cite{Kretschmann W}, however, it allows us to avoid the
technicalities required to analyze a truly, genuinely, infinite
many-body system). Let us consider a specific length of chain $N$,
split into $v=N/(l+s)$ {\it sections}, each consisting of one live
block of length $l$ and one spacer block of length $s := \delta l <<
l$. In the following the sizes $N,l$ will generally be taken to be
large enough that the statements we use hold. The Live channel will
be defined by:
\begin{equation}
{\cal{E}}_{live}: A \rightarrow {\rm tr}_{\rm env}\{U (\rho_{L_1
L_2....L_v} \otimes A) U^\dag\}
\end{equation}
where $A$ represents the state that Alice inputs to the live
channel, $U$ represents the interaction between the environment and
$A$, the labels $L_1,L_2,...,L_v$ represent the live blocks from
sections $1,..,v$, and the trace is taken over the environment. Due
to translational invariance the reduced state of the environment
corresponding to each given live block will be same, and so let us
denote this state by $\rho^l_N$. With this notation, the {\it
product} channel will be defined by:
\begin{equation}
{\cal{E}}_{product}: A \rightarrow {\rm tr}_{\rm env}\{U
((\rho^l_N)^{\otimes v} \otimes A) U^\dag\}.
\end{equation}
Note that both the Live and Product channels have a dependence upon
both the live block length $l$ and the total number of spins $N$.
Let us first consider using the Product and Live channels to send
{\it classical} information. By definition, if a given rate $R$ is
achievable for the Product channel, then for every error tolerance
$\epsilon > 0$ there is an integer $N_\epsilon$ such that for $n >
N_\epsilon$ channel uses there exist a set of $\nu = \lfloor 2^{nlR}
\rfloor$ codeword $nl$-qubit states $\{\rho_1,...,\rho_\nu\}$ and a
corresponding decoding measurement $\{M_1,...,M_{\nu}\}$ such that:
\begin{equation}
\mbox{tr}\{{\cal{E}}_{product}(\rho_i) M_i\} \geq 1 - \epsilon
\,\,\,\, \forall i \in 1...\nu. \label{codecondition}
\end{equation}
If the same codebook and decoding measurements are used without
alteration for the Live channel, then the error would be:
\begin{eqnarray}
\mbox{tr}\{[{\cal{E}}_{live}(\rho_i) - {\cal{E}}_{product}(\rho_i)]
M_i\} \nonumber \\ + \,\, \mbox{tr}\{{\cal{E}}_{product}(\rho_i)
M_i\}. \label{liveerror}
\end{eqnarray}
As the addition of Alice's state $A$, the unitary interaction $U$,
and the POVM element $M_i$ can all be viewed as one new POVM element
acting only on the environment, the left term in this formula can be
bounded by \cite{Nielsen C}
\begin{eqnarray}
|\mbox{tr}\{[{\cal{E}}_{live}(\rho_i) - {\cal{E}}_{product}(\rho_i)]
M_i\}| \nonumber \\ \leq {1 \over 2} ||\rho_{L_1 L_2....L_v} -
(\rho^l_N)^{\otimes v}||_1 \nonumber
\end{eqnarray}
where {$||X||_1:=$tr$\{\sqrt{X^\dag X}\}$} is the {\it trace norm}.
Hence the error (\ref{liveerror}) in using the product code for the
Live channel can be bounded by:
\begin{eqnarray}
\mbox{tr}\{{\cal{E}}_{live}(\rho_i) M_i\} \geq 1 - \epsilon -{1
\over 2} ||\rho_{L_1 L_2....L_v} - (\rho^l_N)^{\otimes v}||_1
\nonumber
\end{eqnarray}
Assume that the rightmost term in this equation is bounded by:
\begin{eqnarray}
||\rho_{L_1 L_2....L_v} - (\rho^l_N)^{\otimes v}||_1 \leq C \, v  \,
l^E \exp (-Fs) \label{decay}
\end{eqnarray}
for positive constants $C,E,F$. This assertion will be demonstrated for
some special cases in section \ref{proof1}.
Then this would mean that the error becomes bounded as
\begin{eqnarray}
\mbox{tr}\{{\cal{E}}_{live}(\rho_i) M_i\} \geq 1 - \epsilon - C \, v
\, l^E \exp (-Fs).
\end{eqnarray}
The $\epsilon$ part of this error depends upon the number of blocks
$v$. One potential problem that we immediately face
is that to decrease $\epsilon$ we need to increase $v$, however,
increasing $v$ inevitably increases the last error term in the equation. It
is hence not a priori clear that both error terms can be made to decrease simultaneously.
However, it can be shown \cite{Kretschmann W,private} that if
we pick $v=l^5,s=\delta l, \delta
>0$ then both error components can be made to vanish as $l$ increases,
while still operating at the achievable rates of the product channels
(in fact, the number of sections $v$ could be given any polynomial
or subexponential dependance on $l$ provided that asymptotically
$v(l) > l^5$).

So we see that provided condition (\ref{decay}) can be demonstrated for
the many-body systems that we consider, then the Product channel works well for
the Live channel, as long as a large enough live block size is used (however small the fraction of spacer qubits
$\delta$). Hence equation (\ref{decay}) is the first of our sufficient conditions. In section \ref{proof1}
we demonstrate that condition (\ref{decay}) (which is identical to equation (\ref{decayrepeat})
later in the paper) holds for some interesting classes of many-body system, including matrix
product states.

\subsubsection{Step [B]}

Now that we know that the product code is also suitable for the live
channel, it is necessary to check that the regularized Holevo bound
(i.e. the regularized Holevo bound for the full channel without
throwing spins away) is actually an achievable rate for the live/spacer blocking
code that has been used. In order to make this analysis it will
be convenient to define a little more notation. For a total chain of
length $n$ as before let ${\cal E}_n$ denote the noisy channel. For
a contiguous subset of $j \leq n$ of the spins that Alice sends, let
${\cal E}^j_n$ denote the effect of the channel only upon those
spins. Due to translational invariance the location of the spins is
irrelevant, as long as they form a contiguous block.

\begin{figure}[t]
\resizebox{8.5cm}{!}{\includegraphics{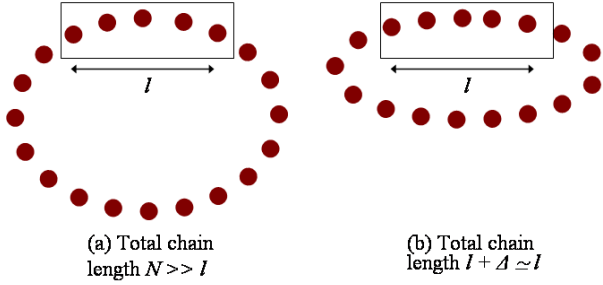}} \caption{
To show that the Product channel (which is just a product of the
reduced channel on a single Live block) Holevo capacity is
essentially the regularized capacity, we need to show that the
reduced channel on a single Live block is essentially independent of
the total length of the chain. Hence we need to show that the
reduced state of $l$ contiguous environment spins is approximately
the same regardless of whether the chain is (a) much longer than
$l$, or (b) slightly longer than $l$.} \label{requirements2}
\end{figure}

A given product channel with live block length $l$ and a total
number of spins $N=v(l+s)=l^6(1+\delta)$ has a Holevo quantity given
by:
\begin{eqnarray}
\chi({\cal E}^l_N) = \chi ({\rm tr}_{\rm env}\{U ((\rho^l_N) \otimes
\bullet ) U^\dag\}),
\end{eqnarray}
where the $\bullet$ merely acts as a place holder for the inputs to
the channel. Our goal is to show that for large enough $l$ this
expression is close to the regularized Holevo bound equation
(\ref{crep}) (see figure \ref{requirements2}).  It is not too
difficult to derive conditions under which this will be the case.
Suppose that we have a spin chain of total length $l+ \Delta (l)$
where $\Delta(l) << l$. In fact we will only be considering
functions $\Delta (l)> 0$ such that $\lim_{l \rightarrow \infty}
{\Delta (l) \over l} = 0$. The {\it subadditivity} and the {\it
Araki-Lieb} inequalities for the entropy (\cite{Nielsen C}, section
11.3.4), i.e.,
\begin{equation}
S(A) + S(B) \geq S(AB) \geq |S(A)-S(B)|
\end{equation}
can be inserted straightforwardly into the Holevo bound to show
that:
\begin{equation}
\chi ({\cal E}^l_{l+\Delta}) \geq \chi ({\cal E}_{l+\Delta}) - 2
\Delta \log (d)
\end{equation}
where $d$ is the dimension of each communication spin (see also \cite{Kretschmann W}). This equation
follows from the fact that the Holevo bound is the difference of two
entropic terms, each of which can change by at most $\Delta \log(d)$
under the tracing out of $\Delta$ $d$-level particles. Dividing
through by $l$ now gives:
\begin{equation}
{\chi ({\cal E}^l_{l+\Delta}) \over l} \geq {l+ \Delta \over l}{\chi
({\cal E}_{l+\Delta}) \over l + \Delta} - 2 {\Delta \over l}\log (d)
\label{holdiff}
\end{equation}
This equation tells us that the Holevo quantity for a subset of $l$
spins is very close to the Holevo quantity for a {\it full} chain of
$l+\Delta$ spins, as long as $\Delta$ is small. Our goal now is to
show that if the subset of $l$ spins is drawn from a much longer
chain of length $N=l^6(1+\delta)$, then the subset still has
essentially the same value for the Holevo quantity, and so the
regularized Holevo quantity represents the capacity of the product
channel. Intuition suggests that if the correlations decay fast
enough, then it should be the case that for $N=l^6(1+\delta)$ we
should have approximately ${\cal E}^l_{l+\Delta}\sim{\cal E}^l_{N}$,
as a given region shouldn't `feel' how long the chain is. Now
suppose that we define
\begin{equation}
P = P(l,\Delta) := ||\rho^l_{l+\Delta} - \rho^l_{N}||_1 =
||\rho^l_{l+\Delta} - \rho^l_{l^6(1+\delta)}||_1.
\end{equation}
Then for a given input state $\omega$ on the live block in question
the output states will differ by at most:
\begin{eqnarray}
||{\rm{tr}_{env}}\{U[\omega \otimes (\rho^l_{l+\Delta} -
\rho^l_{l^6(1+\delta)})]U^\dag\}||_1 \nonumber \\
\leq ||U[\omega \otimes (\rho^l_{l+\Delta} -
\rho^l_{l^6(1+\delta)})]U^\dag||_1 \leq P(l,\Delta).
\end{eqnarray}
Hence Fannes inequality \cite{Fannes 73} (of which a version
suitable for our purposes is $|S(X)-S(Y)|\leq ||X-Y||_1 \log(d) +
\log(e)/e$) can be used to bound the difference in the two Holevo
functions $\chi({\cal E}^l_{l+\Delta}),\chi({\cal E}^l_{N})$ as
follows:
\begin{eqnarray}
{ \chi({\cal E}^l_{l^6(1+\delta)}) \over l} \geq { \chi({\cal
E}^l_{l+\Delta}) \over l} - 2\left(1 \over l \right)\left(P\log
(d^l) + {\log(e) \over e}\right) \nonumber
\end{eqnarray}
Putting this equation together with equation (\ref{holdiff}) gives:
\begin{eqnarray}
{ \chi({\cal E}^l_{l^2(1+\delta)}) \over l} \geq &&{l+ \Delta \over
l}{\chi ({\cal E}_{l+\Delta}) \over l + \Delta} \nonumber \\ && -  2
{\Delta \over l}\log (d) - 2\left(1 \over l \right)\left(P\log (d^l)
+ {\log(e) \over e}\right) \nonumber
\end{eqnarray}
and taking the limit of large $l$ gives:
\begin{eqnarray}
\lim_{l \rightarrow \infty} { \chi({\cal E}^l_{l^2(1+\delta)}) \over
l} \geq \chi_{\infty} - \lim_{l \rightarrow \infty} 2\left(P\log
(d)\right) \nonumber
\end{eqnarray}
So as long as long as we can pick a function $\Delta(l)$ such that
$\lim_{l\rightarrow \infty} \Delta(l)/l = 0$, and such that the norm
distance $P(l,\Delta(l))$ vanishes with increasing $l$ then we know
that the regularized Holevo quantity is the correct capacity.

\subsubsection{Step [C]}

Now that we have understood the conditions under which the
regularized Holevo bound represents the capacity for the
transmission of {\it classical} information, we need to try to
undertake the same analysis for {\it quantum} information. As was
also exploited in \cite{Kretschmann W}, the way that Devetak's work
\cite{Devetak} proves that the regularized coherent information
equals the quantum channel capacity of {\it memoryless} channels is
to first prove a capacity formula for the transmission of {\it
private} (secret) classical information, and then to make the
private coding scheme {\it coherent}. This `coherentification'
procedure applies directly to correlated channels, and so to argue
that the regularized coherent information (\ref{qrep}) is also
achievable for channels with correlated noise, it is sufficient to
show that the private information codes that work for the Product
channel are also suitable for the Live channel. So now suppose that
a malicious eavesdropper is in charge of the environment of our
correlated channel. We need to prove that the information that she
can access is still limited when Product private codes are used for
the Live channel. We can see that the output that Eve obtains is
given by:
\begin{equation}
{\cal{E}}^{\rm Eve}_{live}: A \rightarrow {\rm tr}_{\rm
sys}\{\tilde{U} (\tilde{\rho}_{L_1 L_2....L_v} \otimes A)
\tilde{U}^\dag\}
\end{equation}
where the tildes mean that environment state $\rho$ must be extended
to give a closed system (i.e. $\tilde{\rho}$ is a pure state), the
entire environment of which is assumed to be totally under Eve's
control. In the case of the Product channel the privacy condition
means that for all $\epsilon >0$ there is a $v_0$ such that for all
$v>v_0$ there exists some standard state $\theta$ such that:
\begin{equation}
||{\rm tr}_{\rm sys}\{\tilde{U} ((\tilde{\rho}^l_N )^{\otimes
v}\otimes A) \tilde{U}^\dag\} - \theta||_1 \leq \epsilon
\end{equation}
for all inputs $A$ from the privacy code (readers familiar with
\cite{Devetak,Kretschmann W} will note that in those works an extra
{\it randomisation} index was included as a label in the code states
- however, in our context this is unimportant and so we omit it for
ease of notation). Applying the same code to the Live channel gives
the estimates:
\begin{eqnarray}
&& ||{\rm tr}_{\rm sys}\{\tilde{U} (\tilde{\rho}_{L_1
L_2....L_v}\otimes A) \tilde{U}^\dag\} - \theta||_1 \nonumber
\\ \leq && \epsilon + ||{\rm tr}_{\rm sys}\{\tilde{U} ([\tilde{\rho}_{L_1
L_2....L_v} - (\tilde{\rho}^l_N )^{\otimes v}]\otimes A)
\tilde{U}^\dag\}||_1 \nonumber \\ \leq && \epsilon +
||\tilde{\rho}_{L_1 L_2....L_v} - (\tilde{\rho}^l_N )^{\otimes
v}||_1
\end{eqnarray}
The last term in this equation represents the norm difference
between the {\it purifications} of two different possible
environmental states. We are free to pick the purifications that
give the greatest overlap between the two environment states.
Although this may seem like a contradictory step, as we should allow Eve
to have control over the environment, it is in fact valid because the product
code is by assertion private for {\it all} possible extensions
of the product channel. The
coherentification procedure leads to the distribution of maximally
entangled states which are automatically uncorrelated from the
environment, whatever purification Eve decided to use. The last line from the previous equation hence becomes
(using the fact that for two pure states the overlap and the trace
distance are related by $||(|\phi\rangle \langle \phi| -|\psi\rangle
\langle \psi|)||_1 = 2 \sqrt{1-|\langle \psi|\phi\rangle|^2}$, see
Nielsen \& Chuang \cite{Nielsen C}, p. 415 eq. (9.99), noting that
the factor of 2 comes in from a different convention for the trace
norm):
\begin{eqnarray}
\leq \epsilon + 2 \sqrt{1-F^2({\rho}_{L_1
L_2....L_v},({\rho}^l_N)^{\otimes v})}
\end{eqnarray}
where $F$ is the Uhlmann fidelity \cite{Nielsen C}. Hence, using the
well known relationship between the Uhlmann fidelity and the trace
norm of two states ($1-F(x,y) \leq 2||x-y||_1 \leq
\sqrt{1-F(x,y)^2}$, Nielsen \& Chuang \cite{Nielsen C}, page 416,
from which one can obtain $\sqrt{1-F(x,y)^2} \leq
\sqrt{2(1-F(x,y))}\leq 2\sqrt{||x-y||_1}$), we find that:
\begin{eqnarray}
&& ||{\rm tr}_{\rm sys}\{\tilde{U} (\tilde{\rho}_{L_1
L_2....L_v}\otimes A) \tilde{U}^\dag\} - \theta||_1 \nonumber
\\ \leq && \epsilon + 4\sqrt{||{\rho}_{L_1
L_2....L_v}-({\rho}^l_N)^{\otimes v}||_1}
\end{eqnarray}
Putting the norm bound (\ref{decay}) (which we have not yet
justified) into this equation gives:
\begin{eqnarray}
&& ||{\rm tr}_{\rm sys}\{\tilde{U} (\tilde{\rho}_{L_1
L_2....L_v}\otimes A) \tilde{U}^\dag\} - \theta||_1 \nonumber
\\ \leq && \epsilon + 4\sqrt{C \, v  \,
l^E \exp (-Fs)}
\end{eqnarray}
which is small enough for the assignment $v=l^5$,$s=\delta l$, as
long as $l$ is large enough.

\subsubsection{Summary of sufficient conditions.}

All of this analysis means that in order to argue that the
regularized coherent information and the regularized Holevo bound
are the true quantum or classical capacities, the following two
conditions taken together are sufficient:

(1) To show that the product codes are also good for the partitioned
memory channel, \begin{eqnarray} ||\rho_{L_1 L_2....L_{l^5}} -
(\rho^l_N)^{\otimes l^5}||_1 \leq C \, l^5  \, l^E \exp (-Fs)
\label{decayrepeat}
\end{eqnarray}
for some positive constants $C,E,F$, where $N=l^6(1+\delta)$,$s=\delta l$.

(2) To show that the regularized coherent information is the
appropriate rate for the these codes we need to show that
\begin{equation}
       \lim_{l \rightarrow \infty} ||\rho^l_{l+\Delta(l)} - \rho^l_{l^6(1+\delta)}||_1 =
        \lim_{l \rightarrow \infty} P(l,\Delta(l)) = 0
        \label{longshort}
\end{equation}
for some function $\Delta(l)$ such that $\lim_{l\rightarrow \infty}
\Delta(l)/l = 0$. In fact if equation (\ref{decayrepeat}) holds, in
this condition we could replace $\rho^l_{l^6(1+\delta)}$ with
$\rho^l_{vl(1+\delta)}$ where the number of sections $v$ is any
function of $l$ with a sub-exponential dependance (e.g. a
polynomial) that is asymptotically larger than $l^5$.

To demonstrate that these conditions hold for the most general types
of many-body system is a non-trivial task. However, in a number of
interesting cases it is possible to prove that these conditions
hold. In the remaining sections we demonstrate that these conditions hold
for finitely correlated/matrix product states, as well as for a class of 1D bosonic
system whose ground states may be determined exactly.

\section{Proof of property eq. (\ref{decayrepeat})
for various states}\label{proof1}
In this section we provide proofs for the validity of eq.
(\ref{decayrepeat}) for a variety of quantum states. These
include matrix-product states for which we have discussed
explicit memory channels in this paper. In fact the proofs that
we present for matrix product states are essentially contained in previous works
such as \cite{Fannes NW 92}. We also
demonstrate analogous results for the ground state of quasi-free bosonic systems as
such systems may provide interesting examples for future work. In addition to the
results we present here and in the next section, M. Hastings has demonstrated that
conditions (\ref{decayrepeat},\ref{longshort}) hold for certain
interesting classes of fermionic system \cite{Hastings}.

\medskip

{\em Matrix product or finitely correlated states --} The proof that we present here is essentially one part of the proof of
proposition 3.1 in \cite{Fannes NW 92}. Our presentation of the argument
benefits from the arguments presented in Appendix A of \cite{Wolf VHC 07}
and the review article \cite{PGVWC 06}.

An important tool in the argument is the use of the {\it Jordan canonical form} \cite{Szekeres}.
As some readers may be unfamiliar with this technique, we briefly review it here.
If a square matrix $M$ has complex eigenvalues $\{\lambda_{\alpha}\}$, then it can be shown that a basis may be
found in which the operator can be expressed as the following direct sum:
\begin{equation}
M = \bigoplus_{\alpha}( \lambda_{\alpha} \mathbb{I}_{\alpha} + \mathcal{N}_{\alpha}) \label{jordan}
\end{equation}
where each $\mathbb{I}_{\alpha}$ is an Identity sub-block with an appropriate dimension, and
each $\mathcal{N}_{\alpha}$ is a {\it nil-potent} matrix, meaning that for each $\mathcal{N}_{\alpha}$ there is some
positive integer $k$ such that $\mathcal{N}^{k}_{\alpha}=0$. Moreover, each nilpotent matrix $\mathcal{N}_{\alpha}$
itself may be written as a block-diagonal matrix, where each sub-block is either a zero matrix, or is all zero
except possibly for 1s that may be positioned on the super-diagonal. In other words, each sub-block of a given $\mathcal{N}_{\alpha}$  is
either zero or is of
the form:
\begin{eqnarray} \left(\begin{array}{ccccc}
     0 & 1 & 0 & 0 &. \\
     0 & 0 & 1 & 0 &. \\
     0 & 0 & 0 & 1 &.\\
     0 & 0 & 0 & 0 &. \\
     . & . & . & . &. \end{array}\right) \label{superdiag}.\end{eqnarray}
The decomposition (\ref{jordan}) is the Jordan canonical form of $M$. In our case the matrix
$M$ will be constructed from a completely positive map that can be associated
to the matrix product states that we consider. One consequence of this, for reasons that we discuss
later, is that we will ultimately
only be interested in operators $M$ whose eigenvalues satisfy $1 = \lambda_1 = |\lambda_{1}| > |\lambda_{2}| \geq |\lambda_{3}| \geq ...$.
For a related reason we will also only be interested matrices $M$ for which there is a {\it unique} eigenvector
corresponding to $\lambda_1$, and also for which the sequence of integer powers $M^r$, $r=1,...,\infty$ is bounded.

For matrices obeying these extra conditions we may exploit the Jordan normal form in the following way.
Pick the smallest integer $k$ such that $\mathcal{N}^{k+1}_{\alpha}=0$ for all $\mathcal{N}^{k}_{\alpha}$.
Then $M^r$ can be written as follows:
\begin{equation}
M^r = \bigoplus_{\alpha} \left[ \sum_{m=0,..,k} {r \choose m} \lambda^{r-m}_{\alpha} \mathcal{N}^m_{\alpha}\right].
\end{equation}
If $r$ is large, then all blocks corresponding to $\alpha \neq 1$ will become small because of the
$\lambda^{r-m}_{\alpha}$ term, and so the only sizeable contribution to $M^r$ will come from
the block corresponding to $\alpha=1$, i.e. the sub-block:
\begin{equation}
\left[ \sum_{m=0,..,k} {r \choose m}  \mathcal{N}^m_{1}\right] \label{mainterm}
\end{equation}
Now we have asserted that the sequence of operators $M^r$ is bounded. However, it is not
too difficult to show that for $r=1..\infty$ the sequence of operators (\ref{mainterm}) becomes unbounded
if $\mathcal{N}_{1}$ is non-zero.
This means that if the sequence of operators $M^r$ is bounded, we are forced to conclude
that $\mathcal{N}_{1}=0$, and hence as $M$ has a unique maximal
eigenvector, this means that $\mathbb{I}_{1}$ is an identity
matrix of dimension $1\times1$, i.e. $\mathbb{I}_{1}=1$.

Putting all this together means that a square matrix $M$ with a unique maximal eigenvalue 1, such that the sequence
$M^r$ is bounded, may be
decomposed as:
\begin{equation}
M = 1 \oplus \bigoplus_{\alpha \neq 1} (\lambda_{\alpha} \mathbb{I}_{\alpha} + \mathcal{N}_{\alpha})
\label{specialjordan}
\end{equation}
This means that $M^r$ can be written in the form:
\begin{equation}
M^r = 1 \oplus \lambda^{r}_2 \left[ \bigoplus_{\alpha \neq 1}  \sum_{m=0,..,k} {r \choose m} \left({\lambda^{r-m}_{\alpha}\over \lambda^{r}_2}\right) \mathcal{N}^m_{\alpha} \right]
\label{specialjordan2}
\end{equation}
For our purposes it will be convenient to pull out a factor $r^k$ from the term in square
brackets:
\begin{equation}
M^r = 1 \oplus r^k \lambda^{r}_2 \left[ \bigoplus_{\alpha \neq 1}  \sum_{m=0,..,k} {{r \choose m} \over r^k} \left({\lambda^{r-m}_{\alpha}\over \lambda^{r}_2}\right) \mathcal{N}^m_{\alpha} \right]
\label{specialjordan3}
\end{equation}
This has the advantage of making the operator in square brackets bounded
even as $r \rightarrow \infty$.
This form for $M^r$ will be extremely useful to us. We will apply it to
a completely positive map that can be associated to any matrix product state.
Using this, we will show the decay of correlations required.

The relationship between matrix product states
and CP maps is described in detail in articles such as \cite{Fannes NW 92,PGVWC 06}. Any matrix product state
can be generated by repeatedly acting on a fictitious ancilla particle using an appropriately
constructed CP map. Suppose that we have a matrix product state
 of $N$ particles $j \in \{1,..,N\}$, each associated with
a Hilbert space $H_j$. Consider also a fictitious
`generator' ancilla system on a finite dimensional space $H_{gen}$. It can be shown that the state of the $N$ particles
in the matrix product state can be defined as the state that results from
an appropriate CP map ${\cal T}:{\cal{B}}(H_{gen}) \rightarrow {\cal{B}}(H_{gen})\otimes {\cal{B}}(H_{j})$
which generates each particle $j \in \{1,..,N\}$ in sequence. The generating ancilla
is then traced out to give the matrix product state of the $N$ particles. Related to the map ${\cal T}$ is
the completely positive map ${\cal Q}$, which is the restriction of the map ${\cal T}$ to
the generator ancilla as both input and output. The map ${\cal Q}$ essentially represents
the transfer matrix of the MPS -
for a review of how to construct ${\cal T}$ for matrix product
states, see the article \cite{PGVWC 06}.

The starting state of the fictitious generator ancilla
is usually taken as a fixed point of ${\cal Q}$, in order that the MPS be translationally
invariant. Away from a phase transition point, the CP map ${\cal Q}$ has a unique fixed point of eigenvalue
1, with all other eigenvalues of absolute value strictly less than 1. Let this fixed
point of ${\cal Q}$ be the state $\sigma$. Furthermore,
as ${\cal Q}$ is a CP map, it is clear that the sequence of maps ${\cal Q}^r$ is bounded.
Hence as ${\cal Q}$ acts as a finite dimensional linear operator taking the ancilla space to itself, we
can also think of it as a square matrix and apply equation (\ref{specialjordan3}) to represent powers ${\cal Q}^r$ of the map.
Let us use this form to compute the action of ${\cal Q}^r$ on an input density matrix $\omega$ of the fictitious ancilla. As any
density matrix is taken to a density matrix by a CP map, we may apply (\ref{specialjordan3}) to give that the output
of ${\cal Q}^r$ must have the following form:
\begin{equation}
{\cal Q}^r(\omega) =  \sigma + r^k \lambda^{r}_2 \,\, \Theta_r
\end{equation}
where in the second term $\Theta_r$ is a sequence of operators whose norm can be bounded,
and the $r^k \lambda^{r}_2$ term which governs the size of the deviation from
the final fixed point $\sigma$ arises as a consequence of equation (\ref{specialjordan3}).
This equation essentially states that the deviation of ${\cal Q}^r(\omega)$ from
$\sigma$ falls off as fast as $r^k \lambda^{r}_2$. Although the explicit form
of $\Theta_r$ depends upon the input state, a bound on the norm of $\Theta_r$ can easily
be constructed that is independent of $\omega$. This means that $\lim_{r \rightarrow \infty} {\cal Q}^r = \Sigma$,
where we define $\Sigma$ as the (idempotent) channel that
discards the input ancilla state and creates a copy of $\sigma$ in its place. For finite
$r$ we may write:
\begin{equation}
{\cal Q}^r =  \Sigma + r^k \lambda^{r}_2 \,\, \Theta' \label{qr}
\end{equation}
where $\Theta'$ is now represents operations of bounded norm acting on states of the ancilla
(we have dropped the potential $r$-dependence of $\Theta'$ to keep notation uncluttered,
as it is unimportant).

Our goal in the remainder of this subsection will be to apply this deviation
estimate to show that equation (\ref{decayrepeat}) holds for matrix product systems.
This can be done in two steps. In the first step we show that for two large blocks
of length $L$ separated by a distance $d$ (eventually $L$ will become the length
of the live blocks $l$, and $d$ will become the spacer distance $\delta l$) the
reduced state can be approximated by a product. The second step will use the triangle
inequality to go from this result to the full condition (\ref{decayrepeat}).

The first step proceeds as follows. For convenience we will consider a chain of total chain of length $2n+2L+d$, for
which
the state of the whole chain can be written:
\begin{equation}
{\rm{tr}_{anc}}\{T^{n + L +d + L + n}(\sigma)\}.
\end{equation}
If we take the
limit as $n \rightarrow \infty$, the reduced state of the two large blocks $A,B$ each of length $L$ can be written
\begin{equation}
\rho_{AB}= {\rm{tr}_{anc}}\{\Sigma {\cal T}^L {\cal Q}^d {\cal T}^L \Sigma (\sigma)\},
\end{equation}
and the individual reduced states of each block $A,B$ can be written:
\begin{equation}
\rho_{A}= {\rm{tr}_{anc}}\{\Sigma {\cal T}^L \Sigma (\sigma)\},
\end{equation}
and
\begin{equation}
\rho_{B}= {\rm{tr}_{anc}}\{\Sigma {\cal T}^L \Sigma (\sigma)\}.
\end{equation}
Now from equation (\ref{qr}) we know that up to a correction $d^k \lambda^d_2 \Theta'$, the channel ${\cal Q}^d$ becomes
equivalent to $\Sigma$. Hence we find that $\rho_{AB}$ and $\rho_A \otimes \rho_B$
deviate as follows:
\begin{eqnarray}
\parallel \rho_{AB} - \rho_{A} \otimes \rho_{B}\parallel &=& d^k \lambda^d_2 \parallel {\rm{tr}_{anc}}\{\Sigma  {\cal T}^L  \Theta' {\cal T}^L \Sigma(\omega)\} \parallel \nonumber  \\
&\leq & {\rm constant} \times d^k \lambda^d_2
\end{eqnarray}
where the constant is independent of $L$. Now for our situation $L$ is
simply the size of each block $l$, and the spacing between the blocks
is $s=\delta l$. Hence for two live blocks separated by one spacer block this bound becomes:
\begin{eqnarray}
\parallel \rho_{AB} - \rho_{A} \otimes \rho_{B}\parallel &\leq& {\rm constant} \times s^k \lambda^{s}_2 \nonumber \\
&\leq& {\rm constant} \times s^k \exp(\log(\lambda_2) s) \nonumber
\end{eqnarray}
To go from this result for two live blocks to equation (\ref{decayrepeat})
one simply notes that the above argumentation can also be applied
to blocks of unequal size, and then the triangle inequality applied
to sequences sums of a similar structure to
$||\rho_{L_1L_2L_3L_4} - \rho_{L_1L_2L_3}\otimes \rho_{L_4}|| +
||\rho_{L_1L_2L_3}\otimes \rho_{L_4}- \rho_{L_1L_2}\otimes \rho_{L_3}\otimes \rho_{4}||$
yields equation (\ref{decayrepeat}) with only a polynomial
overhead in $l$.

\medskip

{\em Bosonic systems --} Here we consider chains of harmonic
oscillators whose Hamiltonian can be written in the form
\begin{equation}
    \label{hamiltonian}
    H= {\hat {\bf p}}{\hat {\bf p}}^\text{T} /2 +
    {\hat {\bf x}} V {\hat {\bf x}}^\text{T}/2
\end{equation}
where $\hbar=1$ and we arrange the canonical conjugate position
and momentum operators in vector form
${\hat {\bf x}}=({\hat x}_1,\dots,{\hat x}_{n})$ and
${\hat {\bf p}}=({\hat p}_1,\dots,{\hat p}_{n})$ and introduced
the so-called potential matrix $V$ \cite{Audenaert EPW 02}. The
potential matrix encodes the interaction pattern of the harmonic
oscillators in the chain. From now on we assume that $V$ is a
k-banded matrix, i.e. $V_{i,j}=0$ for $|i-j|\ge k/2$. Physically
this implies that interaction strength vanish strictly beyond the
$(k/2)$-th neighbour. An important quantity in this context is
the {\em symplectic} matrix $\sigma$ which is defined by
$\sigma_{jk} = \langle [{\hat R}_j,{\hat R}_k]\rangle$ where we
denote
${\hat {\bf R}}=({\hat x}_1,\ldots,{\hat x}_n,{\hat p}_1,\ldots,{\hat p}_n)$.

The ground state of the Hamiltonian eq. (\ref{hamiltonian}) is
then a Gaussian state \cite{Eisert P 03,PVreview}
in the sense that its characteristic function $\chi_{\rho}(z) =
tr[{\hat \rho} {\hat W}_z]$ where ${\hat W}_{z}=
e^{iz^T\sigma {\hat R}}$ is the Weyl operator is Gaussian, i.e.
\begin{equation}
    \chi_{\rho}(z) = \chi_{\rho}(0) e^{-\frac{1}{4}z^T\sigma^T\gamma\sigma z + D^T z}
    \label{characteristic}
\end{equation}
where $\gamma_{j,k}= 2\text{Re}[{\hat R}_j {\hat R}_k {\hat \rho}]$
and $D=\sigma tr[{\hat R}{\hat \rho}]$. The density operator may
then be recovered via
\begin{equation}
        {\hat \rho} = \frac{1}{(2\pi)^n}\int d^{2n}z \chi_{\rho}(-z)
        {\hat W}_z.
        \label{rho}
\end{equation}
For the ground state the first moments $D$ due to the reflection
symmetry of the Hamiltonian. Therefore the ground state is fully
characterized by the covariance matrix $\gamma$, which is defined as
$\gamma_{j,k}= 2\text{Re}[ {\hat R}_j {\hat R}_k {\hat \rho} ]$
where we have explicitly used the fact that the first moments
vanish. An explicit computation reveals that the covariance
matrix of the ground state of Hamiltonian eq. (\ref{hamiltonian})
is given by $\gamma= V^{-1/2}\oplus V^{1/2}$ \cite{Audenaert EPW 02}.

For the following proof of eq. (\ref{decayrepeat}) we will bound
the trace norm by the quantum relative entropy using
\cite{Ohya P 93}

{\bf Property 1:} For all density operators ${\hat \sigma},{\hat \rho}$
we have $S({\hat \sigma}||{\hat \rho}) \ge \frac{1}{2}(tr|{\hat \sigma}-{\hat \rho}|_1)^2$
and $S({\hat \sigma}_{AB}||{\hat \sigma}_{A}\otimes {\hat \sigma}_{B})=
S({\hat \sigma}_{A}\otimes{\hat \sigma}_{B})-S({\hat \sigma}_{AB})$.

The entropy of a Gaussian state is determined by the {\em symplectic}
eigenvalues $\{\mu_j\}$ of $\gamma$ that are simply the standard
eigenvalues of the $i\gamma\sigma$. We then find \cite{PVreview}
\begin{eqnarray}
    S({\hat \rho}) &=& \sum_{j=1}^N f(\mu_j)
    \label{entropyI}
\end{eqnarray}
where
\begin{equation}
        \label{func}
       f(x) = \frac{x+1}{2} \log_2
        \frac{x+1}{2} - \frac{x-1}{2} \log_2 \frac{x-1}{2}.
\end{equation}
In the following proof we will need to compute reduced density
matrices. On the level of covariance matrices this is particularly
easy as the covariance matrix of a sub-system $A$ is obtained simply
by removing all entries referring to operators in the complement of
$A$.

Before we proceed to the proof of property eq. (\ref{decayrepeat})
we first derive a useful Lemma that extends Fannes inequality to
Gaussian states. Fannes showed \cite{Fannes 73} that for
$d-$dimensional systems and $\Delta=tr|{\hat \rho}-{\hat \sigma}|\le 1/e$, we find
       $ |S({\hat \rho})-S({\hat \sigma})| \le \Delta\log d
        - \Delta\log\Delta.$
Obviously, in this form the theorem cannot be extended to infinite
dimensional continuous variable systems as this would imply
$d\rightarrow \infty$ which renders the upper bound trivial.
Considering Gaussian states however it is possible to derive a more
useful Fannes-type inequality.

{\bf Lemma I (Bosons):} Given two N-mode Gaussian states ${\hat \rho}_i$
characterized by covariance matrices $\gamma_i$ with symplectic
eigenvalues $\{\mu_i^j\}_{j=1,\ldots,N}$ that satisfy $\max_j
|\mu_1^j-\mu_2^j|\le B$, where $B\approx 0.17623008$ is the nonzero
solution of $(k+2)\log_2(k+2)+k\log_2k=2$, we find
\begin{eqnarray}
        |S({\hat \rho}_1)-S({\hat \rho}_2)| &\le&
        \sum_{j=1}^N -|\mu_1^j-\mu_2^j|\log_2|\mu_1^j-\mu_2^j|
        \nonumber\\
        &\le& \Delta \log_2 N - \Delta\log\Delta
\end{eqnarray}
where $\Delta=\sum_{j=1}^N |\mu_1^j-\mu_2^j|$.

\noindent {\bf Proof --} A Gaussian state is a valid quantum
mechanical state exactly if it satisfies the uncertainty relations
$\gamma+i\sigma \ge 0$. This implies $\mu_i^j\ge 1$ for all $i,j$.
To bound the entropy eq. (\ref{entropyI}) we note that for
$f(x)$ as defined in eq. (\ref{func}) we have
\begin{equation}
        \lim_{x\rightarrow 1} [f(x+k)-f(x)+k\log_2 k] \le 0
        \,\,\mbox{for}\,\, k \le B
\end{equation}
and that $\forall x>1$ and $k>0$ we find
$\frac{d}{dx} [f(x+k)-f(x)+k\log_2 k] \le 0.$ Thus we have
$0\le f(x+k)-f(x)\le -k\log_2 k$ for all $x\ge 1$ and $k\le
B$. Inserting this into the entropy formula eq. (\ref{entropyI}) we
then find the first inequality in Lemma I. The second inequality is
obtained from the fact that the entropy of any probability
distribution with $N$ non-zero probabilities is bounded by $\log_2
N$. This completes the proof.

It is worthwhile noting that an analogous theorem may also
be proven for the fermionic case \cite{fermions}.

{\bf Theorem 1:} In an infinite chain of harmonic
oscillators in its ground state we pick two blocks, each consisting
of $L$ contiguous harmonic oscillators. The two blocks are separated
from each other by $d$ harmonic oscillators. Then we find that
\begin{equation}
        ||\rho_{AB} - \rho_A\otimes\rho_B ||_1 \le C(L) e^{-\alpha d}
\end{equation}
for some polynomial $C(L)$ and constant $\alpha$ independent of $d$.\\

{\bf Proof --} We will proceed using Lemma I to bound the
entropy difference $S({\hat \rho}_{AB}||{\hat \rho}_{A}
\otimes {\hat \rho}_{B})= S({\hat \rho}_{A}\otimes{\hat \rho}_{B})
- S({\hat \rho}_{AB})$. To this end we need to bound the
difference in symplectic eigenvalues of the covariance matrices
corresponding to $\rho_A\otimes\rho_B$ and $\rho_{AB}$.
Property 1 then yields the desired result.

We denote with $\gamma_{ground}$ the ground state of
the complete system and write the covariance matrix of the
two blocks of harmonic oscillators (both of length $L$)
in the $(x_1,p_1,x_2,p_2,...)$ ordering as
\begin{equation}
        \Gamma = \left(\begin{array}{cc} \Gamma_A & \Gamma_{AB}\\
        \Gamma_{AB}^T & \Gamma_B
        \end{array}\right).
\end{equation}
Given that the potential matrix $V$ is banded we know from
\cite{Benzi G 99,Plenio EDC 05,Cramer EPD 06} that the entries of
$\gamma_{ground}$ decrease exponentially in the distance $d$
from the main diagonal. Therefore, the entries of $\Gamma_{AB}$
are exponentially decreasing with distance from the lower left
corner whose entry is of the order $C_1 e^{-\alpha d}$.

We employ Theorem VIII.3.9 of \cite{Bhatia} which states that
\begin{equation}
        |||\lambda_i^{\downarrow}(A)-\lambda_i^{\downarrow}(B)|||
        \le \sqrt{cond(S)cond(T)} |||A-B|||
\end{equation}
for every unitarily invariant norm and where $S$ ($T$) diagonalize
$A$ ($B$) and $cond(S)=||S||\cdot ||S^{-1}||$ is the condition number.
Given that the matrix $i\Gamma\sigma$ can be diagonalized by a
matrix of the form $U\Gamma^{-1/2}$ we find
\begin{eqnarray}
        |||(\mu^i_1)^{\downarrow}-(\mu_2^i)^{\downarrow}|||
        &\le&\\
        &&\hspace*{-2.1cm}(cond(\Gamma)cond(\Gamma_A\oplus\Gamma_B))^{1/4}
        |||\sigma(\Gamma - \Gamma_A\oplus\Gamma_B)|||. \nonumber
\end{eqnarray}
By the pinching inequality for Hermitean matrices \cite{Bhatia}
${\cal C}(A)\prec A$ we find
$cond(\Gamma_A\oplus\Gamma_B),cond(\Gamma)\le
cond(\gamma_{ground})$. For the trace norm we then find
\begin{equation}
        ||(\mu^i_1)^{\downarrow}-(\mu_2^i)^{\downarrow}||_1
        \le 2(cond(\gamma_{ground}))^{1/2} ||\sigma\Gamma_{AB}||_1.
\end{equation}
Then $||\Gamma_{AB}||_1\le 2L||\Gamma_{AB}||_2$ and
$||\sigma||_1=4L$ yield
\begin{equation}
        ||(\mu^i_1)^{\downarrow}-(\mu_2^i)^{\downarrow}||_1
        \le 16\sqrt{cond(\gamma_{ground})}L^2 C_2 e^{-\alpha d}
\end{equation}
for constants $\alpha$ and $C_2$ independent of L. Inserting
this into Lemma I finishes the proof.

As with matrix product states, application of the triangular inequality then yields eq.
(\ref{decayrepeat}).

\section{Proof of property eq. (\ref{longshort})
for various states}\label{proof2}

{\em Matrix product or finitely correlated states --} We consider the same states as in section \ref{proof1} and proceed
similarly. We begin by computing
\begin{eqnarray*}
        \rho^l_{l+\Delta(l)} &=& {\rm tr_{anc}}\{{\cal T}^l {\cal Q}^{\Delta(l)}(\sigma)\} \\
        \rho^l_{l^6(1+\delta)} &=& {\rm tr_{anc}}\{{\cal T}^l {\cal Q}^{l^6(1+\delta)-l}(\sigma)\}
\end{eqnarray*}
Applying equation (\ref{qr}) again we can write the powers of ${\cal Q}$ as
\begin{eqnarray*}
        {\cal Q}^{\Delta(l)}&=& \Sigma + \Delta(l)^k \lambda^{\Delta(l)}_2 \Theta_1\\
        {\cal Q}^{l^6(1+\delta)-l} &=& \Sigma + (l^7)^k \lambda^{l^5}_2 \Theta_2
\end{eqnarray*}
for two bounded operators $\Theta_1,\Theta_2$. In this equation in order to unclutter the notation
we have replaced the first $l^6(1+\delta)-l$ with the weaker estimate $l^7$,
and the second one (in the exponent) by the weaker estimate $l^5$ - in fact their
form is not particularly important for what follows. Putting these
expressions for the powers of ${\cal Q}$ into the expressions for the states, we find that for
sufficiently large $l$:
\begin{eqnarray}
        ||\rho^l_{l+\Delta(l)}-\rho^l_{l^6(1+\delta)}|| &\le&
        {\rm constant} \times \lambda_2^{\Delta(l)}.
\end{eqnarray}
Picking $\Delta(l) = l^{1/2}$, for example, hence allows us to satisfy
all the required conditions.

\medskip

{\em Bosonic systems --} As for condition eq. (\ref{decayrepeat})
we consider the ground state for Hamiltonians that are quadratic
in the canonical coordinates ${\hat x}$ and ${\hat p}$ and
k-banded potential matrices $V$. The ground state is then given
by $\gamma= V^{-1/2}\oplus V^{1/2}$.

Let us now consider $\rho^{(1)}_l:=\rho^l_{l+\Delta(l)}$
with covariance matrix $\gamma_1$ and
$\rho^{(2)}_l:=\rho^l_{l^6(1+\delta)}$
with covariance matrix $\gamma_2$, ie the reduced density
matrices of a block of $l$ spins in a chain of $l+\Delta(l)$
harmonic oscillators (described by covariance matrix $\Gamma_1$)
and in a chain of $l^6(1+\delta)$ harmonic oscillators
(described by covariance matrix $\Gamma_2$) respectively.
Now we will demonstrate that the covariance matrices $\gamma_1$
and $\gamma_2$ converge to each other in the limit $l\rightarrow\infty$.
In the following we will chose, for our convenience, $L$
sufficiently large to ensure that $l+\Delta(l)\le l^6(1+\delta)$.

Given a k-banded potential matrix $V$ let us chose a
number $r=\Delta(l)/k$. Then $V^r$ is $\Delta(l)$-banded.
Denote with $F$ the composition of first applying an analytic
matrix function to a covariance matrix and subsequently
picking the sub-block describing the reduced state of a
contiguous block of $L$ harmonic oscillators. Analogously,
denote with $p_r$ the composition of first applying the
$r^{th}$ matrix power followed by picking a sub-block
as before.

Then we conclude $p_r(\Gamma_1)=p_r(\Gamma_2)$ due to the
k-bandedness of $V$. Furthermore, by Bernsteins theorem
(see \cite{Bernstein} for a short introduction) we then
find
\begin{eqnarray*}
        ||F(\Gamma_1) - F(\Gamma_2)|| \!\!&\le&\!\!
        ||F(\Gamma_1) - p_r(\Gamma_1)|| +
        ||p_r(\Gamma_2)-F(\Gamma_2)||\\
        &\le& \frac{4M(\chi)}{\chi^r(\chi-1)}.
\end{eqnarray*}
Because $\chi>1$ (see \cite{Bernstein}) this tends to zero with
$\Delta(L)\rightarrow\infty$. Choosing $F(A)=A^{1/2}$ and
$F(A)=A^{-1/2}$ allows us to then to conclude that the difference
of the covariance matrices $\gamma_1$ and $\gamma_2$
is bounded by an exponentially decreasing function in $\Delta(L)$.

To continue, we proceed in two steps. First we show that the
above property implies the weak convergence of the two reduced
density matrices. Then we use this to show that this is already
enough to imply the trace norm convergence.\\
{\bf Lemma I:} Given two Gaussian states $\rho^{(1)}_L$
and $\rho^{(2)}_L$ above with vanishing displacement and covariance
matrices $\gamma^{(1)}_L$ and $\gamma^{(2)}_L$ such that
$\lim_{L\rightarrow\infty}||\gamma^{(1)}_L-\gamma^{(2)}_L||=0$
then for any sequence $X_L$, where $||X_L||_{1}\le C$,
with finite rank we have
$\lim_{L\rightarrow\infty}tr[(\rho^{(1)}_L-\rho^{(2)}_L)X_L]=0$.\\
{\bf Proof --} Given that the Hamiltonian of the harmonic
chain is gapped we find that $\gamma^{(i)}_L\ge c\id$ for
some constant $c<1$ independent of L. Then chose
$||\gamma^{(1)}_L-\gamma^{(2)}_L||\le \epsilon < c \le 1$,
$|1-e^{-x}|\le 2|x|$ for $x\le 1$, $|1-e^{-x}|\le e^{|x|}$
for all $x$ and $X_L$ as above. We find
\begin{eqnarray*}
        |tr[(\rho^{(1)}-\rho^{(2)})X_L]| &=&\\
        &&\hspace*{-2.5cm} =\frac{1}{(2\pi)^n}
        |\int d^{2n}z \, tr[W(-z)X_L] (\chi^{(1)}_L(z)-\chi^{(2)}_L(z))|\\
        &&\hspace*{-2.5cm} \le \frac{||X_L||_1}{(2\pi)^n}
        \int d^{2n}z\, e^{-\frac{1}{4}z^T\gamma_L^{(1)} z}|1 -
        e^{-\frac{1}{4}z^T(\gamma_L^{(2)}-\gamma_L^{(1)}) z}|\\
        &&\hspace*{-2.5cm} \le \frac{||X_L||_1}{(2\pi)^n}
        \int_{|z|\le 2\epsilon^{-1/4}}\hspace*{-1.2cm}
        d^{2n}z\, e^{-\frac{1}{4}z^T\gamma_L^{(1)} z}|1 -
        e^{-\frac{1}{4}z^T(\gamma_L^{(2)}-\gamma_L^{(1)}) z}|\\
        &&\hspace*{-2.2cm} + \frac{||X_L||_1}{(2\pi)^n}
        \int_{|z|\ge 2\epsilon^{-1/4}} \hspace*{-1.2cm}
        d^{2n}z\, e^{-\frac{1}{4}z^T\gamma_L^{(1)} z}|1 -
        e^{-\frac{1}{4}z^T(\gamma_L^{(2)}-\gamma_L^{(1)}) z}|\\
        &&\hspace*{-2.5cm} \le \frac{||X_L||_1}{4(2\pi)^n}
        \int_{|z|\le 2\epsilon^{-1/4}}\hspace*{-1.2cm}
        d^{2n}z\, e^{-\frac{1}{4}z^T\gamma_L^{(1)} z}
        |z|^2\epsilon\\
        &&\hspace*{-2.2cm} + \frac{||X_L||_1}{(2\pi)^n}
        \int_{|z|\ge 2\epsilon^{-1/4}} \hspace*{-1.2cm}
        d^{2n}z\, e^{-\frac{1}{4}z^T\gamma_L^{(1)} z}|1 -
        e^{-\frac{1}{4}z^T(\gamma_L^{(2)}-\gamma_L^{(1)}) z}|\\
        &&\hspace*{-2.5cm} \le \frac{||X_L||_1\epsilon^{1/2}}{(2\pi)^n}
        \int_{|z|\le 2\epsilon^{-1/4}}\hspace*{-1.2cm}
        d^{2n}z\, e^{-\frac{1}{4}z^T\gamma_L^{(1)} z}\\
        &&\hspace*{-2.2cm} + \frac{||X_L||_1}{(2\pi)^n}
        \int_{|z|\ge 2\epsilon^{-1/4}} \hspace*{-1.2cm}
        d^{2n}z\, e^{-\frac{1}{4}z^T\gamma_L^{(1)} z}
        e^{\frac{1}{4}\epsilon |z|^2}\\
        &&\hspace*{-2.5cm}\le ||X_L||_1\left(\sqrt{\frac{\epsilon}
        {\det\gamma_L^{(1)}}}
        + O(\epsilon^{1/4} e^{-\epsilon^{-1/2}||(\Gamma_1-\epsilon)||})
        \right)
\end{eqnarray*}
where the last line follows from upper bounds on the
error function. Note that the first term on the right
hand side is proportional to $tr\rho_L^2$ which is
bounded by a constant independent of $L$ because
$-\log_2 tr\rho_L^2\le S(\rho_L)$ and the harmonic
chain Hamiltonian obeys an entropy-area law
\cite{Audenaert EPW 02}. Thus for sufficiently small
$\epsilon$ the right hand side becomes arbitrarily
small. This concludes the proof of Lemma I.

Now we need to prove that weak convergence implies
trace-norm convergence for harmonic chains. The following
proof will use in an essential way the fact that
the ground state of bosonic Hamiltonians that are
quadratic in the canonical operators obey an area law
\cite{Audenaert EPW 02,Plenio EDC 05}.

\ignore{Let us now consider $\rho_L^{(1)}:=\rho^L_{L+\Delta(L)}$ with
covariance matrix $\gamma_1$ and
$\rho_L^{(2)}:=\rho^{(1)}_{L^2(1+\delta)}$
with covariance matrix $\gamma_2$, ie the reduced density
matrices of a block of $L$ spins in a chain of $L+\Delta(L)$
harmonic oscillators (described by covariance matrix $\Gamma_1$)
and in an infinite chain (described by covariance matrix
$\Gamma_2$) respectively. }

{\bf Lemma II:} For the ground state of a bosonic
Hamiltonian $H$ that is quadratic in the canonical
coordinates the limit
$\lim_{L\rightarrow\infty}tr[(\rho^{(1)}_L-\rho^{(2)}_L)X_L]=0$
for any sequence $X_L$, with $||X_L||_1\le K$, with finite rank
already implies trace norm convergence
$\lim_{L\rightarrow\infty}||\rho^{(1)}_L-\rho^{(2)}_L||_1=0$.

{\bf Proof:} Given $0<\epsilon<1$. To begin with we write
\begin{eqnarray}
        ||\rho^{(1)}_L -\rho^{(2)}_L||_1
        &\le& ||\rho^{(1)}_L -P\rho^{(1)}_LP||_1
        + ||P\rho^{(1)}_LP -P\rho^{(2)}_LP||_1\nonumber\\
        && + ||P\rho^{(2)}_LP -\rho^{(2)}_L||_1
\end{eqnarray}
for some $P$ that is yet to be determined. We now would
like to establish the existence of a spectral projection
$P$ of finite rank such that $||\rho^{(i)}_L - P\rho^{(i)}_L P ||_1
<\epsilon$. In other words we aim to project onto the
subspace made up of the eigenvectors corresponding to the
$k_m$ largest eigenvalues of $\rho^{(i)}_L$. We argue that
such a projection $P_{i,L}$ exists for each $\rho^{(i)}_L$.
Then one may project onto the subspace spanned by the subspaces
determined by $P_1$ and $P_2$ which defines $P_L$.
What we need is that $k_m$ is bounded independent of $L$.
To see this, it is important to note that the ground
state of $H$ satisfies an area law,
i.e. in the 1-D setting there is a constant C such that
$S(\rho^{(i)}_L)\le C$ for all $L$. Let us denote by
$\{\lambda_k^{\downarrow}\}_{k=0,\ldots,\infty}$ the
decreasingly ordered eigenvalues of $\rho_{L}^{(1)}$.
Note that for all $k$ we have $\lambda_k^{\downarrow}\le \frac{1}{k}$
by $tr \rho^{(i)}_L = 1$.
Thus we find
\begin{equation}
        C\ge -\sum_{k=k_m}^{\infty} \lambda_k^{\downarrow}
        \log_2\lambda_k^{\downarrow} \ge \log(k_m)\sum_{k=k_m}^{\infty}
        \lambda_k^{\downarrow}.
\end{equation}
Therefore we find for the choice $k_m\ge e^{C/\epsilon}$ that
$\sum_{k=k_m}^{\infty} \lambda_k^{\downarrow}\le \epsilon$
for any choice of $L$.
Thus $P_L$ can be chosen to be a rank $k_m$ projector. Thus
$P_L$ is bounded in the trace norm but the subspace onto
which it projects will generally depend on $L$.
\ignore{Note that it is crucial for this argument that the entropy
is bounded, for a logarithmically diverging entropy as
it occurs for critical systems (e.g. in the field limit)
this argument fails.}
Note further that with the above $P_L$ the weak convergence
$\lim_{L\rightarrow\infty}tr[(\rho^{(1)}_L-\rho^{(2)}_L)X_L]=0$
implies that for sufficiently large $L$ we have
that $||P_L\rho^{(1)}_LP_L-P_L\rho^{(2)}_LP_L||_1<\epsilon$. Thus we
find that for any $\epsilon>0$ and sufficiently large $L$
we have $||\rho^{(1)}_L -\rho^{(2)}_L||_1 \le 3\epsilon$
thus establishing the required trace norm convergence.

\section{Generalisations to other interactions}

It is natural to ask whether the approach that we
have adopted can enable progress to be made for
unitary interactions other than controlled-phase
gates (or their higher dimensional analogues).
Some generalisation are immediate. For instance,
given {\it any} channels that are probabilistic
applications of unitaries, where the unitaries
are controlled on different classical or quantum
basis states of the environment, expression (\ref{key})
can easily be shown to be an explicit lower bound
to the regularized coherent information. Hence if
the environment state has sufficiently decaying
correlations, expression (\ref{key})
will also be a {\it lower} bound to channel capacity.
In a similar manner it is likely that any channel
whose capacity can be bounded by such simple entropic
expressions will benefit from similar insights.

\section{Discussion \& Conclusions}

We have considered models of correlated error inspired
by many-body physics, with the aim of demonstrating
behaviour in the capacity that parallels similar behaviour
in the associated many-body systems. In this context a
number of interesting questions which require further
investigation.

The first of these questions regards our initial motivation -
to find models of correlated error that display interesting
non-analytic behaviour. However, non-analytic behaviour in
many-body systems arises only in the {\it thermodynamic limit},
and so our results unfortunately do not really explain why
the non-analyticities that have been observed in papers such
as \cite{Macchiavello P 02,Macchiavello PV 03, Daems} occur
for {\it finite} truncations of the channel.
Furthermore, the quest for `genuine' non-analyticity is actually
open to some debate - by redefining the parameters defining the
channel, it is always possible to remove any non-analytic behaviour.
However, we hope that our work may help to shed light on
non-analytic behaviour for physically relevant parameter choices
such as magnetic fields and inter-particle couplings \footnote{In
this context it may be important to note that this is also an issue
in the definition of phase transitions. Some definitions of phase
transitions avoid this problem by not relying explicitly on any
parametrization, but instead by relying on the divergence of
correlation functions or the non-uniqueness of a `well defined'
thermal state \cite{Bratelli R,Sachdev}. Such definitions avoid
the problems of defining non-analyticity, and may well have
analogues in correlated error channels.}. In realistic models
of correlated error it is such forms of parametrization that
will probably be most important.

It will also be interesting to see how far the approach
adopted here can be extended to other possible system
environment interactions. The channels that we have investigated
above are all of a very  specific kind - as random unitary channels,
they do not permit quantum information to be transmitted from
one system particle to another via the environment. More general
channels with memory will have this property, and so it will
be interesting to understand what effects this qualitative
difference this can make.

Another open question is whether the conditions (\ref{decayrepeat},
\ref{longshort}) can be established for wider families of many-body
system. In addition to the systems
for which we have demonstrated these conditions, recent work by  M. Hastings \cite{Hastings} demonstrates
that they hold for the ground states
of many fermionic systems too. His approach raises
interesting questions concerning topological invariants which
may have further significance for the problems considered in this paper.

Finally, it is important to note that the connections made
in \cite{Plenio V 07} and this work are actually quite natural
- entropies and correlations have a significant role in
statistical physics, and so quantum channel capacities with
correlated error should have some connection to many-body physics.
However, it would be nice to know if there is a deeper link,
perhaps through a more direct connection between coding theory
and the physics of physical systems such as spin chains.

\section{Acknowledgments}

We are grateful to Chiara Macchiavello, David Gross, Matthew
Hastings, Dennis Kretschmann and Reinhard Werner for helpful
discussions. We must especially thank Dennis Kretschmann for
his patient clarifications of the results of \cite{Kretschmann W}.
We are also grateful to an anonymous referee whose thorough refereeing
greatly helped us to improve the manuscript.
This work was funded by the Royal Commission for the Exhibition
of 1851, the Leverhulme Trust a Royal Society Wolfson Research
Merit Award and is part of the QIP-IRC supported by EPSRC
(GR/S82176/0) as well as the Integrated Project Qubit
Applications (QAP) supported by the IST directorate as
Contract Number 015848'.


\begin{thebibliography}{99}
\bibitem{fermions} Indeed we find
{\bf Lemma II (Fermions):} Given two N-mode Gaussian states $\rho_i$
characterized by fermionic covariance matrices $\Gamma_i$ and
fermionic symplectic eigenvalues $\mu_i^j$ that satisfy
$\max_j |\mu_1^j-\mu_2^j|\le 0.6$ then we find
\begin{eqnarray}
        |S(\rho_1)-S(\rho_2)| &\le& 2\sum_{j=1}^N -|\mu_1^j-\mu_2^j|\log_2|\mu_1^j-\mu_2^j|
        \nonumber\\
        &\le& 2(\Delta \log_2 N - \Delta\log\Delta)
\end{eqnarray}
where $\Delta=\sum_{j=1}^N |\mu_1^j-\mu_2^j|$.

{\bf Proof --} Remember that the fermionic symplectic eigenvalues
$|\mu_i^j|\le 1$ and that the entropy is given by
$S(\rho) = \sum_{j=1}^N e(\nu_j)$ with
$f(x) = -\frac{1+x}{2} \log_2 \frac{1+x}{2} - \frac{1-x}{2}
\log_2 \frac{1-x}{2}$. Straightforward analysis
shows that for $|x-y|\le 0.5$ and $0\le x,y\le 1$ we have $|-x\log_2
x+y\log_2 y|\le -|x-y|\log_2|x-y|$ for all $0\le x,y\le 1$. Thus
for all $x,y\in[0,1]$ we find that
$|f(x)-f(y)| \le 2|x-y|\log_2\frac{|x-y|}{2}$.
Inserting this into the entropy formula yields the first inequality
in Lemma II. The second one follows in the same way as that of Lemma
I.
This completes the proof.\\
\end{thebibliography}
\end{document}